 \documentclass[journal]{IEEEtran}

\ifCLASSINFOpdf

\else

\fi

\usepackage{amssymb}
\usepackage{xcolor}
\usepackage{cite}
\usepackage{multirow}
\usepackage{booktabs}
\usepackage{makecell}
\usepackage{amsmath}
\usepackage{textcomp}
\usepackage{bm}
\usepackage{color}
\usepackage{setspace}
\usepackage{algorithm}
\usepackage{amsmath}
\usepackage{algorithmic}
\usepackage{amsthm,amsmath,amssymb}
\usepackage{mathrsfs}
\usepackage{graphics}
\usepackage{array,diagbox,siunitx}
\usepackage{graphicx}
\usepackage{subfigure}
\usepackage{url}
\usepackage{hyperref}
\usepackage[hyphenbreaks]{breakurl}

\usepackage{epsfig}
\usepackage{amssymb}

\usepackage{algorithm}
\usepackage{algorithmic}
\renewcommand{\algorithmicrequire}{\textbf{Intput:}}

\usepackage{lineno,hyperref}
\modulolinenumbers[5]

\newtheorem{assumption}{Assumption}

\newtheorem{proposition}{Proposition}

\newtheorem{remark}{Remark}

\begin{document}

\title{Privacy-Preserving Task-Oriented Semantic Communications Against Model Inversion Attacks}

\author{Yanhu Wang, ~Shuaishuai Guo, \emph{Senior Member}, \emph{IEEE},~Yiqin Deng, \emph{Member}, \emph{IEEE},\\~Haixia Zhang, \emph{Senior Member}, \emph{IEEE}, ~and Yuguang Fang,~\emph{Fellow},~\emph{IEEE}

\thanks{Yanhu Wang, Shuaishuai Guo, Yiqin Deng, and Haixia Zhang are with School of Control Science and Engineering, Shandong University, Jinan, 250061, China (e-mail: 
yh-wang@mail.sdu.edu.cn, shuaishuai\textunderscore guo@sdu.edu.cn, yiqin.deng@email.sdu.edu.cn, haixia.zhang@s du.edu.cn).

Yuguang Fang is with Department of Computer Science, City University of Hong Kong, Hong Kong, China (e-mail: my.fang@cityu.edu.hk).}
}

\maketitle

\begin{abstract}
Semantic communication has been identified as a core technology for the sixth generation (6G) of wireless networks.  Recently, task-oriented semantic communications have been proposed for low-latency inference with limited bandwidth.  Although transmitting only task-related information does protect a certain level of user privacy, adversaries could apply model inversion techniques to reconstruct the raw data or extract useful information, thereby infringing on users' privacy.  To mitigate privacy infringement, this paper proposes an information bottleneck and adversarial learning (IBAL) approach to protect users' privacy against model inversion attacks.  Specifically, we extract task-relevant features from the input based on the information bottleneck (IB) theory.  To overcome the difficulty in calculating the mutual information in high-dimensional space, we derive a variational upper bound to estimate the true mutual information.  To prevent data reconstruction from task-related features by adversaries, we leverage adversarial learning to train encoder to fool adversaries by maximizing reconstruction distortion. Furthermore, considering the impact of channel variations on privacy-utility trade-off and the difficulty in manually tuning the weights of each loss, we propose an adaptive weight adjustment method.  Numerical results demonstrate that the proposed approaches can effectively protect privacy without significantly affecting task performance and achieve better privacy-utility trade-offs than baseline methods.
\end{abstract}


\IEEEpeerreviewmaketitle

\section{Introduction}


\IEEEPARstart{R}{ecently}, deep learning (DL) has achieved tremendous breakthroughs,
demonstrating exciting performance gain in computer vision, natural language processing, speech recognition, and so on.
Inspired by all these successes, more and more researchers attempt to apply DL to communications.
So far, there exist a large number of papers investigating DL-based joint source-channel coding (JSCC)\cite{kurka2020deepjscc}, channel estimation\cite{Luo2020Channel}, interference alignment\cite{He2017opt}, signal detection\cite{He2020Model}, wireless resource allocation\cite{Shen2021Graph}, etc.
Numerical results demonstrate that
DL technology can effectively improve the performance of communication systems\cite{9450827}.
However, many emerging intelligent applications envisioned to be supported in 5G/6G systems,
such as augmented/virtual reality (AR/VR), autonomous driving, and environmental monitoring,
present serious design challenges to these DL-based communication systems that aim to transmit data reliably.
Taking autonomous driving as an example,
during driving, the in-vehicle sensing system collects large amounts of data,
mainly images, high-definition videos, and point clouds\cite{2021arXiv210900172S}.
If all data is directly transmitted to an edge server on the roadside for inference,
it may cause excessive communication delays when effective decision-making strategies are used to save lives of passengers and other citizens.

To address the above problem,
task-oriented semantic communication has been proposed and recognized as a new communication paradigm for 6G mobile communications\cite{CALVANESESTRINATI2021107930, 10012981, Qin2022Survey}.
This new paradigm has received intensive attention lately,
leading to the implementations of wireless image retrieval at the edge\cite{Jankowski2020,Jankowski2021}, one-device edge inference\cite{Shao2022,Shao2020}, and multi-device cooperative edge inference\cite{2021arXiv210900172S}.
It enables low-latency inference by extracting and transmitting task-relevant information at the transmitter.
By transmitting the task-related features instead of the original data, task-oriented semantic communications can prevent privacy leakage to some extent.
However, adversaries could still apply model inversion techniques\cite{Fredrikson2015,2020Adversarial,3403125} to recover the raw data or extract useful information, thereby infringing on users' privacy.
To our best knowledge,
there hardly exists any work addressing privacy issues in task-oriented semantic communications.

Privacy-preservation has been a research focus in mobile devices and wireless networks\cite{Gai2018Privacy}.
Many methods on privacy-preservation have been proposed in the current literature,
such as \emph{k}-anonymity\cite{Sweeney2002}, \emph{l}-diversity\cite{Machanavajjhala2007}, and \emph{t}-closeness\cite{Li2007}.
However, most of them do not scale well to handle high-dimensional data.
Besides, they cannot be easily adapted to semantic communication systems. There are several reasons for this.
Firstly, most semantic communication systems are based on autoencoder architectures, and training is carried out in
an end-to-end manner. Directly transplanting existing privacy methods into semantic communication systems
may result in the inability to train the semantic communication system. Secondly, it may significantly affect the
performance of semantic communication systems, such as differential privacy methods (DP) \cite{Dwork2006Differential,dwork2006calibrating}, which directly inject
noise into transmission features.
Motivated by the aforementioned issues,
in this paper, we investigate privacy-preserving techniques
for task-oriented semantic communications to prevent privacy
breaches.

In this paper,
we propose IBAL - a scheme for privacy-preserving task-oriented semantic communications.
Since the IB theory \cite{Tishby1999} maximizes the mutual information between the intermediate representation and the data labels
while reducing the amount of information between the intermediate representation and the input,
we leverage IB theory to extract task-related features.
Inspired by the work\cite{2020Adversarial},
we introduce adversarial learning\cite{2014Generative} into task-oriented semantic communications to protect privacy.
That is, we simulate the game between the DNNs of adversaries and the DNNs of the transmitter,
and enable the output features of the transmitter to fool potential adversaries by maximizing the reconstruction distortion.
The reason why we choose to maximize distortion is that the transmitter may not know the specific intentions of adversaries in advance. It is difficult to develop a unified and effective privacy protection criterion considering all the diverse intentions of adversaries. Maximizing distortion in the recovering raw data can protect all kinds of private information to some extent.
Compared with prior works\cite{Jankowski2020,Jankowski2021,Shao2022,Shao2020},
we focus on the privacy issue in task-oriented semantic communications.
Compared with the DP approach,
our approach does not cause a significant negative impact on task performance.
Moreover, considering the impact of channel variations on task inference performance and privacy,
we propose a communication scheme based on IBAL.

Our main contributions in this paper can be summarized as follows.
\begin{itemize}
    \item
    We propose a privacy-preserving task-oriented semantic communication system based on the IB theory and adversarial learning (IBAL).
    The proposed system could extract and transmit task-related features for low-latency edge inference
    while enhancing privacy,
    that is, not transmitting features that can be used to reconstruct the original data by adversaries through the model inversion technique.
    \item
    To avoid intractable direct estimation of the mutual information in high-dimensional space,
    we convert the minimization of IB loss into the minimization of an upper bound of IB loss by variational approximation method.
    \item
    Based on IBAL, we propose a communication scheme suitable for dynamic channel conditions.
    In particular, we treat privacy-preserving task-oriented semantic communications as a multi-objective optimization problem,
    and then solve it with the multiple gradient descent algorithm (MGDA), which avoids manually tuning the weights of each loss.
    \item
    We evaluate the effectiveness of the developed privacy-preserving task-oriented semantic communication schemes on image classification tasks.
    Extensive numerical results demonstrate that, compared with baseline methods, the proposed schemes can effectively protect privacy without significantly affecting task performance.
\end{itemize}

The remainder of this paper is organized as follows.
Section II introduces the existing task-oriented semantic communication system model and model inversion attacks, as well as the performance metrics adopted in the simulation.
Problem formulation is given in Section III,
and IBAL and its extensions are proposed for the problem.
Section IV offers the numerical results.
Section V concludes this paper.

\emph{Notation:} In this paper, $Y$ represents a random variable
and $y$ is the realization of the corresponding random variable.
$H(Y)$ stands for the entropy of $Y$.
$I(Y;Z)$ represents the mutual information between $Y$ and $Z$.
$\mathbb{E}(Y)$ represents the expectation of $Y$.
$KL(\cdot || \cdot)$ stands for the Kullback-Leibler divergence, measuring the difference between two probability distributions.
$||\cdot||_2$ represents the $l_2$ norm of a vector.
Finally, $y\sim\mathcal{N}(\mu,\sigma^2)$ denotes that $y$ follows a Gaussian distribution with mean $\mu$ and covariance $\sigma^2$.

\section{System Model}

In this section, we first describe existing task-oriented semantic communication systems,
and then introduce model inversion technique that can reconstruct the transmitted features as the raw data.
Finally, we discuss the metrics used to evaluate the performance of our proposed privacy-preserving task-oriented semantic communication system.

\begin{figure*}[t]
       \centering
       \includegraphics[width=0.88\linewidth]{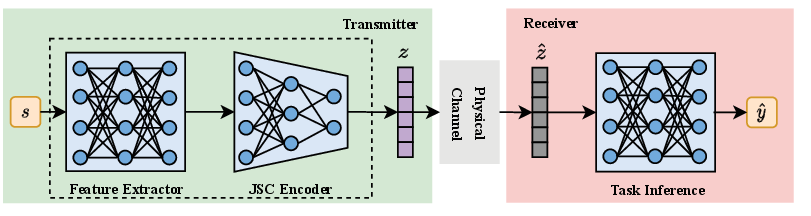}
       \caption{Task-oriented semantic communication system.
       The green region corresponds to the transmitting device, and the red region corresponds to the receiving device.
       The transmitting device extracts and encodes task-related information $z$ from the original data $s$ (e.g., an image),
        and the receiving device directly leverages $\hat{z}$ to obtain the inference result $\hat{y}$.}
       \label{Fig2}
\end{figure*}

\begin{figure*}
       \centering
       \includegraphics[width=0.88\linewidth]{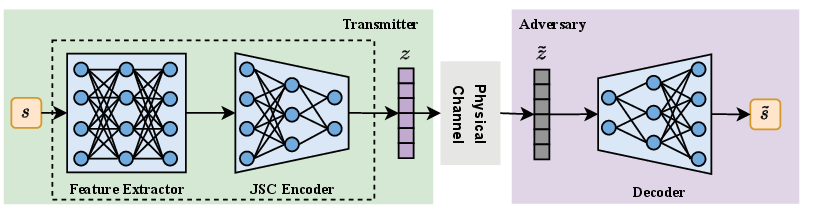}
       \caption{Model inversion attacks.
     The green region corresponds to the transmitting device, and the purple region corresponds to the adversary device.
       The transmitting device transmits the encoded feature $z$.
       Then, an adversary device attempts to produce
       an approximate reconstruction $\tilde{s}$ of the original data based on the illegally obtained $\tilde{z}$, thereby infringing the privacy of the user.}
       \label{Fig3}
\end{figure*}

\subsection{Task-Oriented Semantic Communication Systems}
As shown in Fig. \ref{Fig2}, existing task-oriented semantic communication systems generally consist of a transmitter network, physical channel, and a receiver network,
in which the transmitter network is composed of a feature extractor and a  joint source-channel (JSC) encoder while the receiver network only includes task inference module.

We assume that the system input is $s\in\mathbb{R}^n$, which is from the image space and
contains task-related information and task-irrelevant information.
To enable low-latency inference,
task-related information is extracted by the feature extractor from $s$.
Subsequently, the extracted information is mapped to symbols $z\in \mathbb{C}^{k}$ by the JSC encoder for efficient and reliable transmission.
Note that at the transmitter, the feature extractor can be designed jointly with the JSC encoder,
denoted as $T_{\theta}(\cdot)$, where $\theta$ is the trainable parameter set.
The whole encoding process can be expressed as
\begin{equation}
\label{eq_encoding}
z = T_{\theta}(s),
\end{equation}
The transmitted signal can be attenuated over the channel, and thus the received signal at the receiver can be expressed as
\begin{equation}
\label{eq_channel}
\hat{z} = hz + \epsilon,
\end{equation}
where $\hat{z}\in\mathbb{C}^{k}$,
$h$ represents the coefficient of the Rayleigh fading channel,
and $\epsilon \sim {\cal CN}(0, \sigma^2\mathbf{I}_{k\times k})$.
Finally, the received signal $\hat{z}$ is used by task inference module for edge inference
with the following inference result:
\begin{equation}
\label{eq_decoding}
\hat{y} = R_{\phi}(\hat{z}),
\end{equation}
where $\hat{y}\in \mathbb{R}^l$, and $R_{\phi}(\cdot)$ denotes the task inference network with parameter set $\phi$.

The whole task-oriented semantic communications can be trained by the IB loss function\cite{Shao2022}, which is given by
\begin{equation}
\label{IBLoss}
\mathcal{L}_{IB}= - I(\hat{Z};Y)+\beta I(\hat{Z};S),
\end{equation}
where $S$, $Z$, $\hat{Z}$ and $Y$ are random variables while $s$, $z$, $\hat{z}$, and $y$ are their instances, respectively.
$I(\hat{Z};S)$ is the mutual information, measuring the amount of information between variables $\hat{Z}$ and $S$.
$I(\hat{Z};Y)$ measures the amount of information between $Y$ and $\hat{Z}$.
\subsection{Model Inversion Attacks}
Model inversion attacks, proposed by Fredrikson\cite{Fredrikson2015}, are attacks
in which the adversaries reconstruct the received features as the raw input using DNNs, and then obtain users' privacy.
Below, we provide a detailed description of model inversion attacks.

As shown in Fig. \ref{Fig3},
the left is the transmitter network of the task-oriented semantic communication system, which is actually deployed on the user's device.
The right is the DNN $D_{\gamma}$, designed by the adversary and deployed on the adversary's device.
The encoded feature $z$ of the transmitter is illegally obtained by the adversary, and the obtained feature $\tilde{z}\in\mathbb{C}^{k}$ can be expressed as
\begin{equation}\label{adchannel}
\tilde{z} = h_az + \epsilon_a,
\end{equation}
where $h_a$ represents the adversary channel coefficients, and $\epsilon_a \sim {\cal CN}(\mathbf{0}, \sigma^2_a\mathbf{I}_{k\times k})$.
Then, the adversary attempts to produce an approximate reconstruction $\tilde{s}$ of the user data based on $\tilde{z}$, i.e.
\begin{equation}
\tilde{s} = D_{\gamma}(\tilde{z}).
\end{equation}
To obtain a better reconstruction, the adversary first trains $D_{\gamma}$.
We assume that the adversary performs model inversion through black-box attack\cite{Yuan2019,mohaghegh2020advflow}.
In the process, the adversary can continuously access the network model on the target device, but the adversary cannot obtain the model parameters and training data on the target device.
In detail, the adversary first inputs the image data $s\in \mathcal{S}_A$ into the transmitter network, and then the encoded feature $\hat{z}$ can
be obtained.
Here, we use $\mathcal{S}_A$ to distinguish the adversary's own dataset from the user's private data $\mathcal{S}$. 
After obtaining a large number of data pairs $\{ \left(s\in \mathcal{S}_A, \tilde{z}\right)\}$, the adversary trains its own decoder $D_{\gamma}$ based on these data pairs.
$D_{\gamma}$ is committed to minimize the reconstruction distortion,
thus, the loss function is the mean squared error (MSE):

\begin{equation}  
\label{MSELoss}
\mathcal{L}_{MSE}=\frac{1}{M}\sum_{i=1}^M\left\|s_i-\tilde{s}_i\right\|^2,
\end{equation}
where $s_i$ is the $i$-th sample in $M$ image samples, and $\tilde{s}_i$ is its reconstruction.

\subsection{Performance Metrics}
In this paper, we mainly focus on privacy issues in task-oriented semantic communications
and aim to make the privacy-utility trade-off.
We assume that the image classification task is performed at the receiver,
therefore, the classification accuracy is used to measure the task performance (corresponding to the ``utility").
Classification accuracy can be expressed as
\begin{equation}
\label{accuracy}
{\rm Classification \ accuracy} = \frac{N_{t}}{N_{o}},
\end{equation}
where $N_{t}$ is the number of correctly classified images, $N_{o}$ is the number of classified images.

Through model inversion techniques, adversaries can recover the raw data or extract useful information, thereby infringing on users' privacy.
We maximize the distortion of reconstructed images to reduce the invasion of users' privacy by adversaries.
That is, the higher the distortion of the reconstructed image, the more difficult it is for adversaries to obtain private information from it.
Therefore, we use the quality of images reconstructed by adversaries to measure the level of privacy protection.
Specifically, we use the peak signal-to-noise ratio (PSNR), which is one of the commonly used metrics for measuring image quality. It can be expressed as
\begin{equation}
\label{PSNR}
{\rm PSNR} = 10\log_{10}\frac{\rm MaxValue^2}{\rm MSE} \quad (dB),
\end{equation}
where MSE is the mean squared error,
MaxValue is the maximum value that a pixel can take,
for example, 8-bit image is $2^8-1=255$.
The lower PSNR, the better privacy.

PSNR is a metric of image quality based on the errors between corresponding pixels,
and does not take into account the visual characteristics of human eyes.
The structural similarity (SSIM)\cite{1284395} measures the similarity between images based on their luminance, contrast, and structure,
which better reflects the perceptual characteristics of human vision.
SSIM is given by
\begin{equation}
\begin{split}
\label{ssim}
{\rm SSIM}(s, \hat{s})&= {\rm luminance}(s, \tilde{s})^{\alpha1} + {\rm contrast}(s, \tilde{s})^{\alpha2}  \\&\quad+ {\rm structure}(s, \tilde{s})^{\alpha3}\\
&=\left(\frac{2\mu_s\mu_{\tilde{s}}+c_1}{\mu_s^2 + \mu_{\tilde{s}}^2 + c_1} \right)^{\alpha1} + \left(\frac{2\sigma_s\sigma_{\tilde{s}}+c_2}{\sigma_s^2 + \sigma_{\tilde{s}}^2 + c_2} \right)^{\alpha2}  \\&\quad+\left(\frac{cov(s, \tilde{s})+c_3}{\sigma_s\sigma_{\tilde{s}} + c_3} \right)^{\alpha3},
\end{split}
\end{equation}
where $\mu_s$ and $\mu_{\tilde{s}}$ are the mean values of $s$ and $\tilde{s}$,
$\sigma_s$ and $\sigma_{\tilde{s}}$ are the standard deviations of $s$ and $\tilde{s}$, respectively.
$cov(s, \tilde{s})$ is the covariance between $s$ and $\tilde{s}$.
$\alpha_1$, $\alpha_2$, and $\alpha_3$ respectively represent the proportion of the luminance, the contrast, and the structure in the SSIM,
and their values are typically set to 1.
$c_1$, $c_2$, and $c_3$ are typically set to 0.0001 to avoid division by zero.

\section{Privacy-Preserving Task-Oriented Semantic Communications}

This paper focuses on privacy issues in task-oriented semantic communications.
More specifically, it aims to achieve low-latency edge inference
while preventing transmitted features from being reconstructed for raw data by adversaries through model inversion attacks.
To this end, combining \eqref{IBLoss} and \eqref{MSELoss}, we formulate the optimization problem as
\begin{equation}
\begin{split}
\label{obj}
\min \limits_{\theta, \phi} \quad \lambda \mathcal{L}_{IB}-(1-\lambda)\mathcal{L}_{MSE},\\
\end{split}
\end{equation}
where the range of $\lambda$ is $[0,1]$.
The smaller $\lambda$, the stronger the privacy-preservation.

\subsection{Problem Statement}
Intuitively, it is promising to use \eqref{obj} as the objective function for privacy-preserving task-oriented semantic communications.
The first term of \eqref{obj} is the loss function of the task inference performance,
and the second one could captures the reconstruction capability of the transmitted signals by the adversary.
However, in practice, there are still many issues to be resolved as discussed below.
\begin{itemize}
    \item 
 \textbf{Problem 1}: How to compute $\mathcal{L}_{IB}$?

The mutual information term in \eqref{obj} (i.e., $\mathcal{L}_{IB}$) can be expressed as follows:
\begin{equation}
\begin{split}
\label{LIBZK}
\mathcal{L}_{IB}&= - I(\hat{Z};Y)+\beta I(\hat{Z};S)\\
&= -\int p(y,\hat{z})\log \frac{p(y,\hat{z})}{p(y)p(\hat{z})} \,{dyd\hat{z}}  \\&\quad+\beta \int p(s,\hat{z})\log \frac{p(s,\hat{z})}{p(s)p(\hat{z})} \,{dsd\hat{z}}\\
&= -\int p(y|\hat{z})p(\hat{z})\log p(y|\hat{z}) \,{dyd\hat{z}} - H(Y) \\&\quad+\beta \int p_{\theta}(\hat{z}|s)p(s)\log \frac{p_{\theta}(\hat{z}|s)}{p(\hat{z})} \,{dsd\hat{z}},
\end{split}
\end{equation}
where the entropy $H(Y)$ depends on input data and can be ignored when performing optimization in \eqref{obj}.
As can be seen from the system shown in Fig. \ref{Fig2} and the description in Section II.A,
$p_{\theta}(\hat{z}|s)$ can be expressed as $p_{\theta}(\hat{z}|s) = p_{\theta}(z|s)p_{channel}(\hat{z}|z)$,
where $p_{\theta}(z|s)$ is the encoding distribution
and $p_{channel}(\hat{z}|z)$ represents the channel model.
With $p(\hat{z})$ and $p(y|\hat{z})$ following Markov chain $Y \leftrightarrow S \leftrightarrow \hat{Z}$,
they can be written as
\begin{equation}
\begin{split}
\label{pz}
p(\hat{z})=\int p(s)p_{\theta}(\hat{z}|s)\,{ds},
\end{split}
\end{equation}
and
\begin{equation}
\begin{split}
\label{pyz}
p(y|\hat{z})&=\int p(s,y|\hat{z}) \,{ds}
\\&=\int p(y|s)p(s|\hat{z}) \,{ds}
\\&=\int{ds}\,\frac{p(y|s)p_{\theta}(\hat{z}|s)p(s)}{p(\hat{z})}.
\end{split}
\end{equation}
In \eqref{pz} and \eqref{pyz}, the calculations of integrals are involved,
which is highly complicated in general,
especially in high-dimensional spaces,
which also makes $\mathcal{L}_{IB}$ hard to estimate and optimize.
Thus, to make \eqref{obj} practical in task-oriented semantic communication systems to solve the privacy problem,
it is necessary to develop effective method to estimate $\mathcal{L}_{IB}$.

 \item 
 \textbf{Problem 2}: How to train the whole system?

The existing task-oriented semantic communication system shown in Fig. \ref{Fig2} cannot be trained with \eqref{obj}
because the framework consists of feature extractor, JSC encoder and task inference module,
while \eqref{obj} involves image reconstruction.
Therefore, it is desirable to design a new task-oriented semantic communication system framework,
in which the transmitter network is co-trained with the task inference module while coping with the decoder.
Moreover, the extracted and transmitted features by the transmitter network
should be designed to be hard to decode by any decoder of adversaries other than the intended decoder.
So in theory, all possible decoders should be used in the new communication system framework for the transmitter network to learn.
However, this is not obviously realistic.
Thus, to better protect privacy,
an effective method should be developed for the transmitter to learn to cope with all possible decoders of adversaries,
so that the newly designed task-oriented semantic communication system is not too complicated.

 \item 
 \textbf{Problem 3:} How to adaptively adjust $\lambda$ according to the dynamic channel conditions?

The impact of channel variations on privacy need to be elaborated in more detail.
The adversary reconstructs the original images from the received features
through model inversion to reveal the privacy of a user.
The worse the image reconstruction quality, the better the privacy protection.
When channel conditions from the transmitter to the adversary become worse,
the transmitted features are more severely corrupted
and more difficult to recover from the raw input,
and hence the privacy is enhanced.
Of course, the task inference performance will be affected as well.
In practice, in face of worse (or better) channel conditions,
by considering both the requirements of task inference performance and the enhancement of privacy with channel noise,
the user may increase (or decrease) $\lambda$ in \eqref{obj}.
However, due to the trade-off between privacy and task inference performance,
which we call the privacy-utility trade-off,
it is hard to manually tune $\lambda$ to achieve satisfactory results.
Therefore, when considering the privacy-utility trade-off and the effect of channel variations,
it is essential to design a method
that can better achieve the privacy-utility trade-off.
\end{itemize}

\begin{figure*}
       \centering
       \includegraphics[width=0.85\linewidth,height=0.33\linewidth]{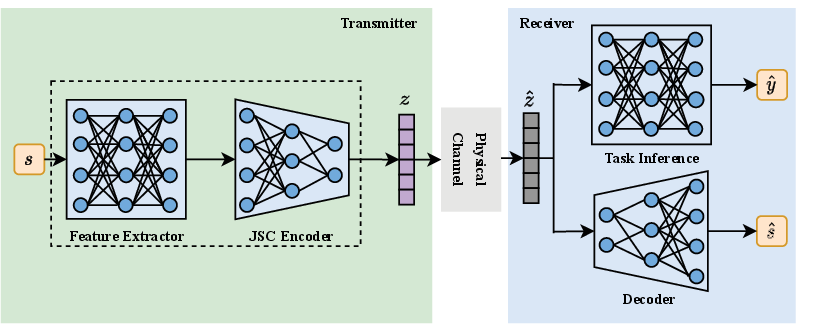}
       \caption{The framework of proposed IBAL.
       The green region corresponds to the transmitting device, and the blue region corresponds to the receiving device.
       Training procedure: We update the decoder using dataset $\mathcal{S}_2$ while fixing the transmitter network and task inference module.
       Then, we update the transmitter network and task inference module using dataset $\mathcal{S}_1$ while fixing the decoder.
       Communication procedure: The transmitting device transmits the encoded feature $z$ that fools the adversary, i.e., prevent reconstructing the input by the adversary, while achieving edge inference through the task inference module.
       }
       \label{Fig4}
\end{figure*}

\subsection{Variational Information Bottleneck}
To solve \textbf{Problem 1},
we leverage the variational method to build a tractable upper bound of information bottleneck loss. 
Specifically,
due to the fact that $p(\hat{z})$ and $p(y|\hat{z})$ in $\mathcal{L}_{IB}$ is hard to compute, we seek a mutual information estimator to compute $\mathcal{L}_{IB}$.
Recently, some works\cite{poole2019variational,alemi2016deep,cheng2020club} propose to combine the variational approximation with deep learning to facilitate the estimation of mutual information.
Following these works, we leverage variational approximation to address our problem at hand, i.e,
\begin{equation}
\begin{split}
\label{qz}
p(\hat{z})=q(\hat{z}) \quad {\rm and} \quad p(y|\hat{z})&= q_{\phi}(y|\hat{z}),
\end{split}
\end{equation}
where $\phi$ is the parameter set of task inference network at the receiver.

By introducing variational approximations \eqref{qz},
we could derive a tractable variational upper bound for $\mathcal{L}_{IB}$ (see \textbf{Appendix A}),
which is given as\cite{Shao2022}:
\begin{equation}
\begin{split}
\label{VIBLoss}
\mathcal{\overline{L}}_{VIB} &=\mathbb{E}_{p(s,y)}\{\mathbb{E}_{p_{\theta}(\hat{z}|s)}[- \log q_{\phi}(y|\hat{z})] \\&\quad+\beta KL(p_{\theta}(\hat{z}|s)||q(\hat{z}))\}.\\
\end{split}
\end{equation}

To obtain an unbiased estimate for $\mathcal{\overline{L}}_{VIB}$,
we utilize the reparameterization trick\cite{Kingma2014a} and Monte Carlo approximation.
In particular, following \cite{poole2019variational},
we use $\frac{1}{M-1}\sum_{j\neq m}p_{\theta}(\hat{z}|x_{j})$ to approximate $q(\hat{z})$,
and then get
\begin{equation}
\begin{split}
\label{RVIBLoss}
\mathcal{\overline{L}}_{VIB}&\approx \frac{1}{M}\sum_{m=1}^M\Bigg \{-\frac{1}{N}\sum_{n=1}^N\log q_{\phi}(y_m|\hat{z}_{m,n}) \\&\quad+\beta \log \frac{p_{\theta}(\hat{z}_m|s_m)}{\frac{1}{M-1}\sum_{j\neq m}p_{\theta}(\hat{z}_m|s_j)} \Bigg \},
\end{split}
\end{equation}
where $M$ represents the number of sample pairs $\{(s_m,y_m)\}_{m=1}^M$,
and $N$ indicates the number of times to sample the channel for each pair $(s_m,y_m)$.
$p_{\theta}(\cdot)$ and $q_{\phi}(\cdot)$ are the encoding network at the transmitter and task inference network at the receiver, respectively.

Finally, formulation \eqref{obj} can then be rewritten as
\begin{equation}
\begin{split}
\label{obj2}
\min \limits_{\theta, \phi} \quad \lambda \mathcal{\overline{L}}_{VIB}-(1-\lambda)\mathcal{L}_{MSE}.\\
\end{split}
\end{equation}

\subsection{The Proposed IBAL}
To solve \textbf{Problem 2}, we propose an information bottleneck and adversarial learning (IBAL) approach.
\subsubsection{Model Description}
The designed IBAL is as shown in Fig.~\ref{Fig4}.
Similar to the transmitter network in the task-oriented semantic communication system shown in Fig. \ref{Fig2},
the transmitter network in IBAL also consists of a feature extractor
and a JSC encoder to map the extracted feature values to channel input symbols for subsequent transmissions.
The difference is that
the feature extractor in IBAL needs to extract task-related features from raw data to ensure edge inference performance,
while ensuring that the extracted features can fool the adversary.

The channel is modeled as a non-trainable layer,
since the gradients need to be back-propagated.
Since we focus on privacy-preserving task-oriented semantic communications,
in this paper, we only consider additive white Gaussian noise (AWGN) channels and Rayleigh fading channels for simplicity.
Besides, considering output power of the transmitter,
an average power constraint is imposed on the transmitted signal vectors,
which is expressed as
\begin{equation}
\label{apc}
\frac{1}{k}\sum_{i=1}^k z_i^2 \leq 1,
\end{equation}
where $k$ represents the dimension of $z$.

In practice, the adversary is not fixed, and it is difficult and impractical to obtain the adversary's network architecture.
We construct a decoder to simulate the adversary and introduce the designed communication system.
The fundamental purpose is to enable the transmitter network to learn to extract and transmit features, which are difficult for the true adversary's DNNs to decode through model inversion.

\subsubsection{Model Training}

To defend against model inversion attacks from any adversary, we utilize adversarial learning in training a semantic communication system as listed in Algorithm \ref{IBAL}.
Specifically, the algorithm describes a game between an adversary trying to recover the original data from the received signal
while a user aiming to achieve privacy-preserving task-oriented semantic communications.
As illustrated in Algorithm \ref{IBAL}, the training process of the IBAL consists of two stages, which are performed alternately.
After initializing the weights $W$ and biases $b$, input the training data $s_1 \in \mathcal{S}_1$ and $s_2 \in \mathcal{S}_2$.
Note that, the algorithm inputs different training data because it simulates a game between an adversary and a user,
and it is impossible for an adversary to have a user's dataset.
The first stage is to simulate an adversary.
We train the decoder by minimizing the MSE loss $\mathcal{L}_{MSE}$.
In this stage, the parameters of the transmitter network will not be updated since the transmitter network is deployed on the device of the user.
The second stage is to train the proposed communication system with \eqref{obj2} as the training objective function.
Different from the first stage, in the second stage, except the decoder, the parameters of other neural networks are updated.
The two phases are performed alternately until the termination criteria is met.

\renewcommand{\algorithmicrequire}{\textbf{Initialization:}}
\begin{algorithm} [htb]
\caption{IBAL Training Algorithm}\label{IBAL}
\label{IBAL}
    \begin{algorithmic}[1]
        \REQUIRE Initialize weights \emph{W} and biases \emph{b}.
        \STATE \textbf{Input:} Input data $s_1 \in \mathcal{S}_1$ and $s_2 \in \mathcal{S}_2$.
        \STATE \textbf{while} termination criteria is not met \textbf{do}
        \STATE \quad \textbf{Simulate an adversary:}
        \STATE \quad \quad $T_{\theta}(s_2) \rightarrow z_2$.
         \STATE \quad \quad Transmit $z_2$ over the channel.
        \STATE \quad \quad $D_{\kappa}(\hat{z}_2) \rightarrow \hat{s}_2$.
        \STATE \quad \quad Calculate the loss $\mathcal{L}_{MSE}$ by \eqref{MSELoss}.
        \STATE \quad \quad Only update parameters of the decoder, $\kappa$.
        \STATE \quad \textbf{Learn the privacy-preserving task-oriented semantic}\\
        \quad \textbf{communication system:}
        \STATE \quad \quad $T_{\theta}(s_1) \rightarrow z_1$.
         \STATE \quad \quad Transmit $z_1$ over the channel.
        \STATE \quad \quad $R_{\phi}(\hat{z}_1)\rightarrow \hat{y}_1$ and $D_{\kappa}(\hat{z}_1) \rightarrow \hat{s}_1$.
        \STATE \quad \quad Calculate the whole loss by \eqref{obj2}.
         \STATE \quad \quad Update parameters of the transmitter network \\
         \quad \quad and task inference module, $\theta$ and $\phi$.
         \STATE \textbf{end while}
        \STATE \textbf{Output:} IBAL network $T_{\theta}(\cdot)$, $R_{\phi}(\cdot)$, $D_{\kappa}(\cdot)$.
     \end{algorithmic}
\end{algorithm}

\subsection{The Proposed IBAL-D}
To solve \textbf{Problem 3}, we extend the proposed IBAL under dynamic channel conditions, namely IBAL-D.

\subsubsection{Method Description}
As described in \textbf{Problem 3},
channel conditions can affect privacy. To capture the impact,
we modify the original objective function in \eqref{obj2}.
We use the noise level $\sigma^2$ to represent the channel condition.
Considering that the image reconstruction quality is inversely proportional to noise,
we introduce $\sigma^2$ into \eqref{obj2} in the form of a decreasing function.
In this paper, for demonstration purpose, we adopt a heuristic discreasing function ${1}/{(1 + \sigma^2)}$ and reformulate the objective function as
\begin{equation}
\begin{split}
\label{obj3}
\min \limits_{\theta, \phi} \quad \lambda \mathcal{\overline{L}}_{VIB}-( 1- \lambda)\frac{1}{1 + \sigma^2}\mathcal{L}_{MSE}.\\
\end{split}
\end{equation}

Next, we need to address how to adaptively adjust the value of $\lambda$ according to the channel condition.
We treat the considered privacy-preserving task-oriented semantic communications as a special multi-objective optimization problem.
For multi-objective optimization,
its fundamental purpose is to achieve Pareto efficiency.
Under Pareto efficiency,
neither task inference performance nor privacy can be optimized without harming the other.
To solve the problem,
we utilize the multiple gradient descent algorithm (MGDA) proposed in \cite{JA2012Multiple}. Specifically, the MGDA performs Pareto optimization based on the Karush-Kuhn-Tucker (KKT) conditions,
which are given as follows:
\begin{equation}
\label{KKT}
\left\{
        \begin{array}{lr}
        \lambda \bigtriangledown_{\theta} \mathcal{\overline{L}}_{VIB} - (1 - \lambda) \bigtriangledown_{\theta} \left(\frac{1}{1 + \sigma^2} \mathcal{L}_{MSE}\right)=0,&\\
        0\leq \lambda \leq1.&\\
        \end{array}
\right.
\end{equation}
The KKT conditions are necessary conditions,
and the solution satisfying \eqref{KKT} is called a Pareto stationary point.
To utilize the MGDA,
we formulate the optimization problem:
\begin{equation}
\begin{split}
\label{obj5}
\min \limits_{\lambda \in [0,1]}
\left\|\lambda \bigtriangledown_{\theta} \mathcal{\overline{L}}_{VIB} - (1 - \lambda) \bigtriangledown_{\theta} \left(\frac{1}{1 + \sigma^2} \mathcal{L}_{MSE}\right)\right\| ^2_2.\\
\end{split}
\end{equation}
At this time, achieving Pareto efficiency becomes solving \eqref{obj5} for $\lambda$,
thereby adaptively adjusting $\lambda$.

According to \cite{JA2012Multiple} and \cite{3326943},
the solution of \eqref{obj5} is either 0 and the result point satisfies \eqref{KKT},
or the solution provides the gradient descent direction of the improved edge inference task.
From \cite{3042817}, it is known that \eqref{obj5} is a convex quadratic optimization problem with linear constraints.
We use the Frank-Wolfe algorithm developed in \cite{3042817} to obtain the optimal $\lambda^*$ (see \textbf{Appendix B} for details).

\subsubsection{Training Procedure}

The training process of the communication network suitable for different channel conditions is illustrated in Algorithm \ref{IBAL-D}.
Similar to Algorithm \ref{IBAL}, the whole process is still performed alternately between a simulating an adversary and a learning user for privacy-preserving task-oriented semantic communications.
The difference mainly lies in the phase for learning privacy-preserving task-oriented semantic communications.
\renewcommand{\algorithmicrequire}{\textbf{Initialization:}}
\begin{algorithm} [htb]
\caption{IBAL-D Training Algorithm}
\label{IBAL-D}
    \begin{algorithmic}[1]
        \REQUIRE Initialize weights \emph{W} and biases \emph{b}.
        \STATE \textbf{Input:} Input data $s_1 \in \mathcal{S}_1$ and $s_2 \in \mathcal{S}_2$.
        \STATE \textbf{while} termination criteria is not met \textbf{do}
        \STATE \quad \textbf{Simulate an adversary:}
        \STATE \quad \quad $T_{\theta}(s_2) \rightarrow z_2$.
         \STATE \quad \quad Transmit $z_1$ over the channel.
        \STATE \quad \quad $D_{\kappa}(\hat{z}_2) \rightarrow \hat{s}_2$.
        \STATE \quad \quad Calculate the loss $\mathcal{L}_{MSE}$ by \eqref{MSELoss}.
        \STATE \quad \quad Only update parameter set of the decoder, $\kappa$.
        \STATE \quad \textbf{Learn the privacy-preserving task-oriented semantic}\\
        \quad \textbf{communication system:}
        \STATE \quad \quad $T_{\theta}(s_1) \rightarrow z_1$.
         \STATE \quad \quad Transmit $z_1$ over the channel.
         \STATE \quad \quad $R_{\phi}(\hat{z}_1)\rightarrow \hat{y}_1$ and $D_{\kappa}(\hat{z}_1) \rightarrow \hat{s}_2$.
         \STATE \quad \quad Calculate $\mathcal{\overline{L}}_{VIB}$ and update $\phi$. Calculate $\mathcal{L}_{MSE}$ but\\
         \quad  \quad not update $\kappa$.
         \STATE \quad \quad Solve \eqref{obj5} to obtain $\lambda^*$.
         \STATE \quad \quad $T_{\theta}(s_1) \rightarrow z_1$.
         \STATE \quad \quad Transmit $z_1$ over the channel.
         \STATE \quad \quad $R_{\phi}(\hat{z}_1)\rightarrow \hat{y}_1$ and $D_{\kappa}(\hat{z}_1) \rightarrow \hat{s}_1$.
         \STATE \quad \quad Calculate the whole loss by \eqref{obj2} and update $\theta$: $\theta=$\\
          \quad \quad $ \theta -\eta[\lambda^* \bigtriangledown_{\theta} \mathcal{\overline{L}}_{VIB} - (1 - \lambda^*) \bigtriangledown_{\theta}(\frac{1}{1 + \sigma^2} \mathcal{L}_{MSE})]$.
        \STATE \textbf{end while}
        \STATE \textbf{Output:} IBAL-D network $T_{\theta}(\cdot)$, $R_{\phi}(\cdot)$, $D_{\kappa}(\cdot)$.\\
        \quad
     \end{algorithmic}
\end{algorithm}
\subsection{Convergence and Computational Complexity Analysis}
This subsection first analyzes the convergence of our algorithm.
The Adam optimizer is adopted for the proposed IBAL approach due to its outstanding convergence speed and performance.
Thus, the convergence analysis is related to the Adam optimizer.
For simplicity, the objective function is denoted as $F(\textbf{\emph{w}})$, where $\textbf{\emph{w}}= \{ \theta, \phi \}$.
Before analyzing convergence, we present several assumptions commonly made in the literature.

\begin{assumption}
  $F(\textbf{w})$ has an L-Lipschitz gradient in the sense that there exists $L>0$, such that for any $\textbf{w}$ and $\textbf{w}'$ satisfying:
  \begin{equation}
    \begin{split}
    \label{LS}
    F(\textbf{w}')-F(\textbf{w}) \leq \langle \nabla F(\textbf{w}), \textbf{w}'-\textbf{w}  \rangle +\frac{L}{2}\left\| \textbf{w}'-\textbf{w} \right\|_2^2
    \end{split}
  \end{equation}
\end{assumption}

\begin{assumption}
  The gradient is bounded in the sense that $ \mathbf {g}_t = \nabla F_t\left(\textbf{w}^{(t)}\right)$ has an upper bound, i.e.
  \begin{equation}
    \label{GB}
    \left\{
        \begin{array}{lr}
        \left\| \nabla F\left(\textbf{w}^{(t)} \right) \right\|_2 \leq G, \forall t, &\\
        \left\| \mathbf {g}_t \right\|_2 \leq G, \forall t.&\\
        \end{array}
    \right.
\end{equation}
\end{assumption}

\begin{assumption}
The noisy gradient is unbiased and the noise is independent, i.e., $\textbf{g}_t =  \nabla F\left(\textbf{w}^{(t)}\right) + \textbf{n}_t$, $\textbf{n}_t$ satisfies $\mathbb{E}\left[ \textbf{n}_t\right]=\textbf{0}$,
and $\textbf{n}_{t_1}$ is independent of $\textbf{n}_{t_2}$ when $t_1 \neq t_2$.
\end{assumption}

\begin{assumption}
  $\mathbf {g}_1 = \nabla F_1\left(\textbf{w}^{(t)}\right)$ has a lower bound, i.e.,
  \begin{equation}
    \begin{split}
    \label{1B}
    \left\| \mathbf {g}_1 \right\|_2 \geq c, \forall t
  \end{split}
  \end{equation}
\end{assumption}

\begin{remark}
  Assumptions 1-3 are the commonly used assumptions in analyzing convergence \cite{Reddi2018iclr,Chen2019iclr}.
  For Assumption 4, it refers to the scenario that the algorithm not converging at the initial value, which is easily satisfied.
\end{remark}

Our algorithm converges because of the following proposition.

\begin{proposition}
Supposing that the assumptions 1-4 hold, with $\beta_{1,t} \leq \beta_1 \in [0,1)$ and $\beta_{1,t}$ is non-increasing, $\eta_t = \frac{1}{\sqrt{t}}$, for any number of iterations $T$, we have
\begin{equation}
\begin{split}
   \label{slmb}
    {\rm min}_{t=1,2,\ldots,T} \mathbb{E} \left[ \left\| \nabla F \left(\textbf{w}^{(t)}\right) \right\|^2 \right] = \frac{1}{\sqrt{T}} \left(G_1+G_2 \log T\right),
\end{split}
\end{equation}
where $G_1$ and $G_2$ are two constants that are independent of $T$. It can be observed that the convergence rate of our proposed algorithm is $\mathcal{O}\left(\frac{\log T}{\sqrt{T}}\right)$.
\end{proposition}

\emph{Proof:} See Appendix C.

The computational complexity of the proposed IBAL mainly comes from convolution operations, because they involve multiplications and additions. The computational cost of a convolution layer can be expressed as $F \times F \times C_{in} \times C_{out} \times W_1 \times W_2$, where $F \times F$ is the size of the filter, $C_{out}$ is the number of filters, $C_{in}$ represents the number of input channels, and $W_1 \times W_2$ is the size of the feature map.
Assuming that the network has $K$ convolution layers, the computational complexity of the proposed IBAL is $\mathcal{O}\left(\sum\limits_{i=1}\limits^K(F^{(i)})^2 C_{in}^{(i)}  C_{out}^{(i)} W_1^{(i)} W_2^{(i)}\right)$.

\section{Experiments and Discussions}

 In this section, we compare the proposed privacy-preserving task-oriented semantic communications schemes with other DL-based communication schemes
under static and dynamic AWGN and Rayleigh fading channels.
Image classification task is chosen as an example to demonstrate the key idea of the proposed approaches, which is to develop an end-to-end learning framework to extract privacy-preserving task-oriented channel-robust low-dimensional latent representations of data for transmission.
It is noteworthy that the proposed approaches are not specified for the image classification task.

\begin{table}[htb]\scriptsize
\caption{The architecture settings of each module for MNIST classification task}
\label{mniststructure}
\centering
\renewcommand{\arraystretch}{1.5}
\begin{tabular}{c|c|c}
\hline
&  \textbf{Layer name}                                                          & \begin{tabular}[c]{@{}c@{}}\textbf{Output}\\\textbf{dimensions}\end{tabular}\\
\hline
\textbf{Transmitter}                                               & Dense+Tanh                                                                 & 64                                                  \\
\hline
\begin{tabular}[c]{@{}c@{}}\textbf{Receiver}\\\textbf{(Classifier)}\end{tabular} & \begin{tabular}[c]{@{}c@{}}Dense+ReLU\\Dense+ReLU\\Dense+Softmax\end{tabular} & \begin{tabular}[c]{@{}c@{}}1024\\256\\10\end{tabular}  \\
\hline
\begin{tabular}[c]{@{}c@{}}\textbf{Receiver}\\\textbf{(Decoder)}\end{tabular} & Dense+Tanh                                                                 & 784\\
\hline
\end{tabular}
\end{table}

\begin{table}[htb]\scriptsize
\caption{The architecture settings of each module for CIFAR-10 classification task}
\label{cifarstructure}
\centering
\renewcommand{\arraystretch}{1.5}
\begin{tabular}{c|c|c}
\hline
&  \textbf{Layer name}                                                          & \begin{tabular}[c]{@{}c@{}}\textbf{Output}\\\textbf{dimensions}\end{tabular}\\
\hline
\textbf{Transmitter}                                              & \begin{tabular}[c]{@{}c@{}} (Convolutional layer+ReLU)$\times$2\\Residual Block\\(Convolutional layer+ReLU)$\times$3\\Reshape+Dense+Tanh\end{tabular}                                                     & \begin{tabular}[c]{@{}c@{}}128$\times$16$\times$16\\128$\times$16$\times$16\\8$\times$4$\times$4\\128\end{tabular} \\
\hline
\begin{tabular}[c]{@{}c@{}}\textbf{Receiver}\\\textbf{(Classifier)}\end{tabular} & \begin{tabular}[c]{@{}c@{}}Dense+ReLU+Reshape\\Convolutional layer+ReLU\\Residual Block\\Pooling\\Dense+Softmax\end{tabular} & \begin{tabular}[c]{@{}c@{}}8$\times$4$\times$4\\512$\times$4$\times$4\\512$\times$4$\times$4\\512\\10\end{tabular}  \\
\hline
\begin{tabular}[c]{@{}c@{}}\textbf{Receiver}\\\textbf{(Decoder)}\end{tabular} & \begin{tabular}[c]{@{}c@{}}Dense+Tanh+Reshape\\Convolutional layer+ReLU\\(Deconvolutional layer+ReLU)$\times$2\\Residual Block\\Deconvolutional layer+ReLU\\Convolutional layer+Sigmoid\end{tabular}                                                     & \begin{tabular}[c]{@{}c@{}}8$\times$4$\times$4\\512$\times$4$\times$4\\128$\times$16$\times$16\\128$\times$16$\times$16\\64$\times$32$\times$32\\3$\times$32$\times$32\end{tabular} \\
\hline
\end{tabular}
\end{table}
\begin{table}[t]\scriptsize
\caption{The network architecture built by the adversary in the model inversion attack}
\centering
\label{miattack}
\renewcommand{\arraystretch}{1.5}
\begin{tabular}{c|c|c}
\hline
&  \textbf{Layer name}                                                          & \begin{tabular}[c]{@{}c@{}}\textbf{Output}\\\textbf{dimensions}\end{tabular}\\
\hline
\textbf{MNIST}                                               & Dense+Tanh                                                                 & 784                            \\
\hline
\begin{tabular}[c]{@{}c@{}}\textbf{CIFAR-10}\end{tabular} & \begin{tabular}[c]{@{}c@{}}Dense+Tanh+Reshape\\Convolutional layer+ReLU\\Convolutional layer+ReLU+Upsampling\\Residual Block\\Convolutional layer+ReLU+Upsampling\\Residual Block\\Convolutional layer+ReLU+Upsampling\\Residual Block\\Convolutional layer+Sigmoid\end{tabular} & \begin{tabular}[c]{@{}c@{}}8$\times$4$\times$4\\512$\times$4$\times$4\\256$\times$8$\times$8\\256$\times$8$\times$8\\128$\times$16$\times$16\\128$\times$16$\times$16\\64$\times$32$\times$32\\64$\times$32$\times$32\\3$\times$32$\times$32\end{tabular}  \\
\hline
\end{tabular}
\end{table}

\subsection{Experiment Setup}
\subsubsection{Datasets}
To demonstrate the performance of the proposed schemes,
the benchmark datasets MNIST and CIFAR-10 are adopted.
The MNIST dataset contains 70,000 grayscale images of handwritten digits with a resolution of $28 \times 28$,
of which 60,000 are training images and 10,000 are testing images.
The CIFAR-10 is a color image dataset with 10 classes.
There are 6000 images in each class,
including 5000 training images and 1000 testing images.
The resolution of each image is $32 \times 32$.
Note that for training images in the MNIST dataset,
we further split them into two parts,
$\mathcal{S}_2$ with 2/3 for updating $T_{\theta}$ and $R_{\phi}$,
$\mathcal{S}_1$ with 1/3 for updating $D_{\kappa}$.
The same goes for the CIFAR-10 dataset.

\subsubsection{Neural Network Architecture}

\begin{figure*}[!t]
\centering
\vspace{-0.35cm}
\subfigtopskip=2pt
\subfigbottomskip=2pt
\subfigcapskip=-2pt
\subfigure[AWGN]{
\centering \includegraphics[width=0.45\linewidth, keepaspectratio=false]{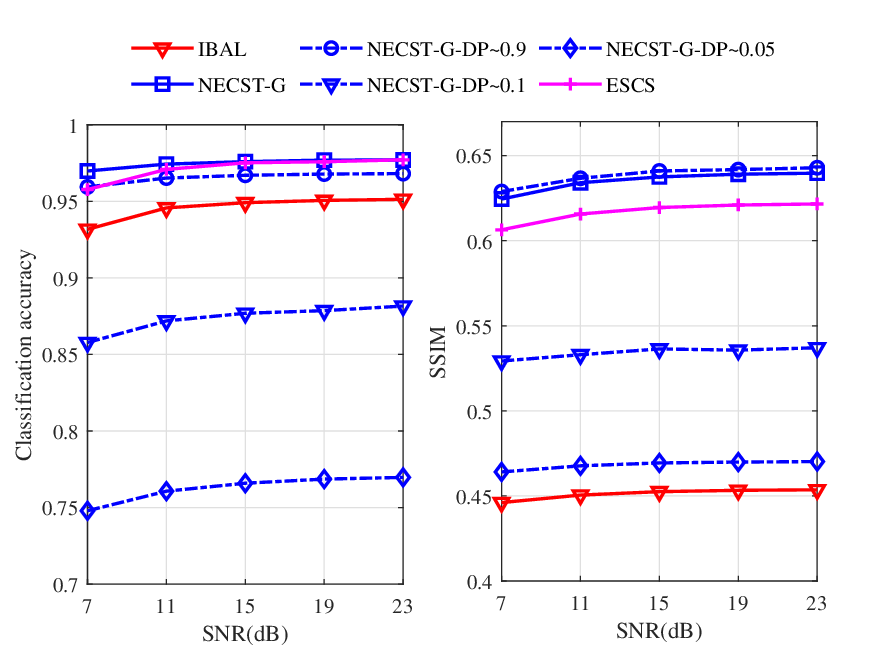}
}
\
\subfigure[Rayleigh Fading]{
\centering \includegraphics[width=0.45\linewidth, keepaspectratio=false]{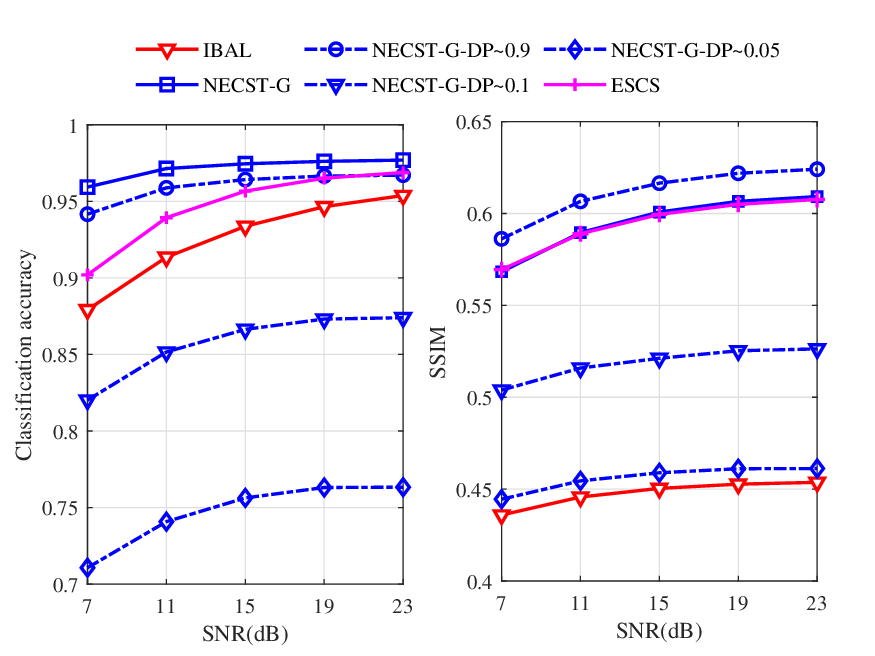}
}
\caption{Classification accuracy and SSIM comparison of IBAL with baseline schemes on the MNIST dataset over (a) AWGN and (b) Rayleigh fading channels.}
\label{MNIST_AR}
\end{figure*}

\begin{figure*}[!t]
       \centering
       \includegraphics[width=0.85\linewidth]{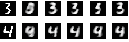}
       \caption{MNIST: from left to right, original, IBAL, NECST-G, NECST-G-DP$\sim$0.9, NECST-G-DP$\sim$0.1, NECST-G-DP$\sim$0.05 and ESCS.}
       \label{MNIST34}
\end{figure*}
In the experiments,
we design different neural network structures for the MNIST and CIFAR-10 datasets as there is still no universal encoder/decoder network for different datasets. The detailed architecture settings can be found in Table \ref{mniststructure} and Table \ref{cifarstructure}.
During the training process, we adopt the Adam optimizer with $\beta_1 = 0.9$, $\beta_2 = 0.99$,
and learning rate $lr=0.001$.

To verify the privacy-preserving capabilities of several baselines and the proposed scheme,
we simulate the adversary using model inversion attack to reconstruct input images using the features transmitted through all communication schemes.
Table \ref{miattack} shows the network architecture built by the adversary in the model inversion attack.
Besides, all baselines adopt the same classification network as the proposed schemes for fair comparison.

\subsubsection{Comparison algorithms}

This paper selects the following DL-based communication schemes as baselines.
\begin{itemize}
\item
 \textbf{NECST-G:} NECST\cite{2018NECST} is a DL-based JSCC scheme, where the encoder outputs are processed with discrete Bernoulli distributions and then pass through the binary erasure channel (BEC) and the binary symmetric channel (BSC).
In the experiments, we process the outputs with Gaussian distributions and directly transmit the encoder output through the AWGN channel and the Rayleigh channel.
To distinguish the original NECST, the adopted NECST is denoted as NECST-G.
We use the cross-entropy as the loss function of NECST-G.
\item
  \textbf{NECST-G-DP:} DP\cite{Dwork2006Differential}  offers a strong mathematical foundation for privacy preservation. 
  By adding Laplace noise within the DP framework, the privacy level can be precisely controlled.
  In the experiments, we will combine NECST-G with DP.
  Specifically, we inject Laplace noise into the transmitted features.
  In the experiments, we will combine NECST-G with DP.
  Specifically, we inject Laplace noise into the transmitted features.
  In the following experiments, we set the location and scale parameters of Laplace distribution to $0$ and the reciprocal of the privacy budget, respectively.
    The privacy budget is set to $0.05$, $0.1$ and $0.9$, respectively.
  A smaller privacy budget means more noise injected into the transmitted features, which also implies stronger privacy protection.
\item
  \textbf{ESCS\cite{Luo2023Encrypted}:} ESCS is an encrypted semantic communication system for privacy preserving, which utilizes symmetric encryption algorithms. Both the encryptor and decryptor are composed of neural networks, and according to the setup in \cite{Luo2023Encrypted}, we use a key with one token in the experiment.
  The loss function of ESCS is set to the cross-entropy.
\end{itemize}

\subsection{Experiment Under Static Channel Conditions}

Firstly, we investigate the performance of the proposed IBAL under static channel conditions.
The performance of the proposed IBAL is compared with ESCS, NECST-G, NECST-G-DP$\sim$ $0.9$, NECST-G-DP$\sim$$0.1$, and NECST-G-DP$\sim$$0.05$.
The proposed IBAL and comparison schemes are trained under SNR=15dB and tested under 7dB, 11dB, 15dB, 19dB and 23dB, respectively.
In the proposed IBAL, $\lambda$ is set to 0.5.
In the experiment, we simulate the adversary using features transmitted through all communication schemes to reconstruct input images,
and calculate the SSIM and PSNR between the reconstructed and original images to characterize the privacy protection capability of communication schemes.
We also use the classification accuracy to characterize edge inference performance.

Fig. \ref{MNIST_AR} shows the classification accuracy versus SNR and SSIM versus SNR for various schemes under AWGN and Rayleigh fading channels on the MNIST dataset.
As can be seen from Fig. \ref{MNIST_AR}(a), compared with NECST-G, the classification accuracy of IBAL is slightly decreased, but the privacy protection ability is significantly enhanced.
When NECST-G uses DP to enhance privacy, edge inference performance is more seriously affected.
For example, NECST-G-DP$\sim$0.05 provides the similar privacy protection to IBAL, but the classification accuracy is $0.15\sim0.2$ lower than IBAL.
The classification accuracy of ESCS is higher than that of IBAL, but its privacy protection capability is significantly weaker than IBAL, and its performance is roughly similar to NECST-G-DP$\sim$0.9.
Fig. \ref{MNIST_AR}(b) shows the same tendency.
Fig. \ref{MNIST34} intuitively shows the MNSIT images obtained after reconstructing the transmitted signals of each communication scheme using model inversion attacks.
These imply the necessity of our study on privacy-preserving task-oriented semantic communications.
It can be found that NECST-G-DP$\sim$0.05 is comparable to the proposed IBAL in privacy preservation, where the former injects Laplace noise into NECST with a privacy budget pf 0.05.
In DP, the smaller the privacy budget, the more noise is being injected.
Due to the injection of more Laplacian noise, the privacy protection capability of NECST-G-DP$\sim$0.05 is enhanced, but its classification accuracy is greatly reduced.
On the contrary, the proposed IBAL obtains the smallest SSIM with acceptable task performance loss.
\begin{figure}[t]
\vspace{-0.5cm}  
    \centering
\includegraphics[width=0.95\linewidth]{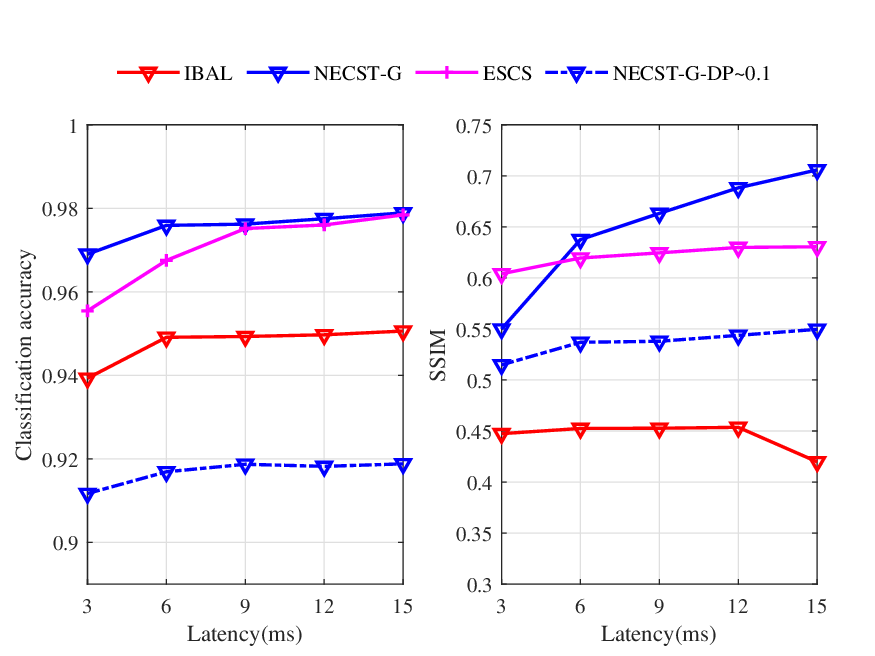}
\caption{Classification accuracy and SSIM of communication schemes in the MNIST classification task under different latency with SNR=15dB.}
\label{SY}
\end{figure}


In addition, we assume that the wireless channel models have the same SNR value 15dB in the training and testing phases.
Then, we record the inference accuracy as well as the level of privacy protection for different communication latency.
In the experiments, the bandwidth is set to $12.5$kHz with a symbol rate of $9,600$ Baud, corresponding to a limited bandwidth.
Fig. \ref{SY} shows the classification accuracy and SSIM values of IBAL, ESCS, NECST-G, and NECST-G-DP$\sim$0.1 on the MNIST dataset.
It is observed that, under the same latency, compared with other algorithms, our proposed IBAL can not only provide the best privacy protection, but also ensure good task inference performance.
This shows that our proposed method can extract task-related features efficiently, and these features are difficult to decode by the adversary's decoder.

\begin{figure}[t]
\centering
\vspace{-0.35cm}

\centering \includegraphics[width=0.9\linewidth, keepaspectratio=false]{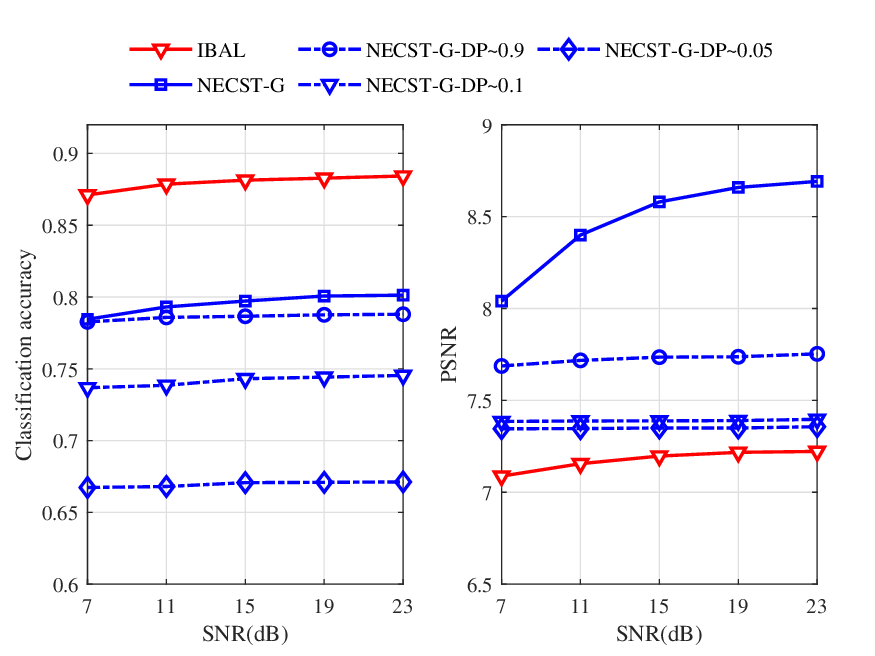}
\
\caption{Classification accuracy and PSNR comparison of IBAL with baseline schemes on the CIFAR-10 dataset over AWGN channels.}
\label{cifar_ar}
\end{figure}

\begin{figure*}
       \centering
       \includegraphics[width=0.86\linewidth]{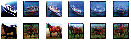}
       \caption{CIFAR-10: from left to right, original, IBAL, NECST-G, NECST-G-DP$\sim$0.9, NECST-G-DP$\sim$0.1 and NECST-G-DP$\sim$0.05.}
       \label{CHM}
\end{figure*}

Next, we also investigate the classification accuracy and PSNR obtained by the proposed IBAL and NECST-G, and NECST-G-DP on the CIFAR-10 dataset, as illustrated in Fig. \ref{cifar_ar}.
It can be seen that the proposed IBAL not only achieves the best classification accuracy but also obtains the smallest PSNR.
In Fig. \ref{CHM}, the CIFAR-10 images obtained after reconstructing the transmitted signals of these schemes using model inversion attacks are shown.
These results further validate that the proposed IBAL improves the privacy-preserving ability without significantly affecting the task inference performance,
which achieves better privacy-utility trade-off compared to baselines.

\begin{figure}
       \centering
       \includegraphics[width=0.95\linewidth]{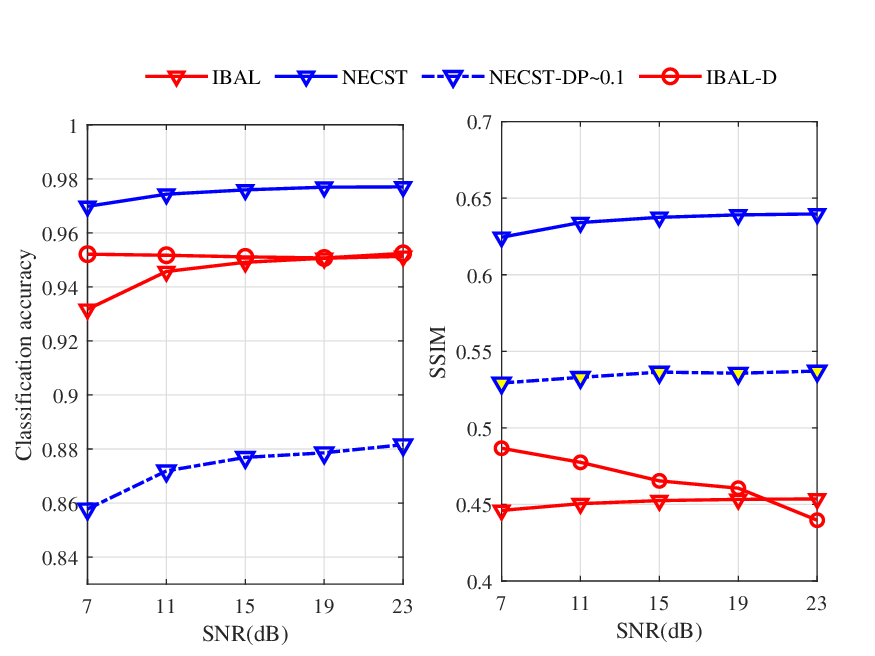}
       \caption{Classification accuracy and SSIM comparison of IBAL-D with other schemes on the MNIST dataset over AWGN channel.}
       \label{dtt}
\end{figure}


\subsection{Experiment Under Dynamic Channel Conditions}

The main purpose of this experiment is to evaluate the performance of the proposed IBAL-D under dynamic channel conditions.
 We use NECST-G, NECST-G-DP$\sim$0.1, and IBAL as comparison algorithms for IBAL-D.
In the proposed IBAL, $\lambda$ is set to 0.5.
All comparison algorithms are evaluated under SNR=15dB.
For IBAL-D, we let the SNR vary between 7dB and 23dB during training.

Fig. \ref{dtt} shows the classification accuracy and SSIM obtained by the proposed IBAL-D and the comparison algorithms on the MNIST dataset.
It can be observed that compared with the baselines and IBAL,
IBAL-D could ensure a smaller reduction in classification accuracy
while achieving privacy protection.
The proposed IBAL-D can adjust the level of privacy protection according to the channel state, and better guarantee the edge task inference under low SNR.
Specifically, when the channel state is poor,
IBAL-D will increase $\lambda$,
and hence the extracted features can better maintain the inference performance.
The poor channel state can make up for the decrease in the privacy protection ability caused by the increase of $\lambda$.
When the channel state is good,
IBAL-D will decrease $\lambda$, making it more difficult for the transmitted features to be restored as the original data, hence enhancing privacy protection.

\begin{figure}[t]
    \centering
\includegraphics[width=0.95\linewidth]{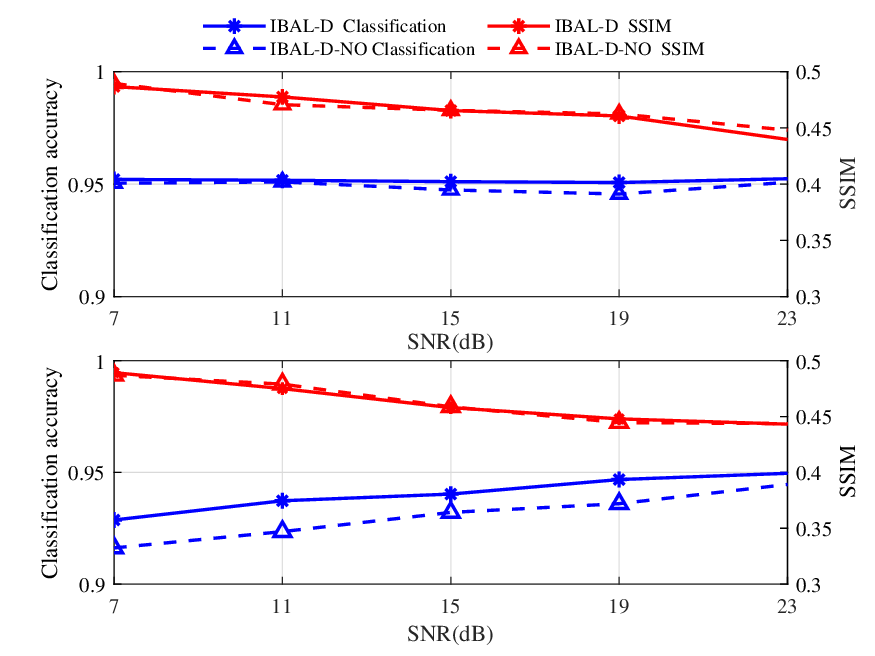}
\caption{Classification accuracy and SSIM comparison of IBAL-D with IBAL-D-NO on the MNIST dataset over AWGN channel (top) and Rayleigh fading channel (bottom).}
    \label{bld}
\end{figure}

In Section III, we treat the privacy-preserving task-oriented semantic communications problem as a multi-objective optimization problem,
and then solve it with the MGDA algorithm.
To verify the effectiveness of the proposed method,
we choose IBAL-D without using the method, namely IBAL-D-NO, as the comparison algorithm.
To ensure fairness,
the SNR during training was also set to vary from 7dB to 23dB for IBAL-D-NO.
Fig. \ref{bld} shows the classification accuracy and SSIM obtained
by the proposed IBAL-D and IBAL-D-NO under AWGN
and Rayleigh fading channels on the MNIST dataset.
It can be seen that with the help of the proposed method,
IBAL-D obtained better privacy-utility trade-off
while avoiding manually adjusting the weights.



\subsection{Performance over Rapidly Changing Channel Conditions}
\begin{figure}[!t]
    \centering
\includegraphics[width=0.95\linewidth]{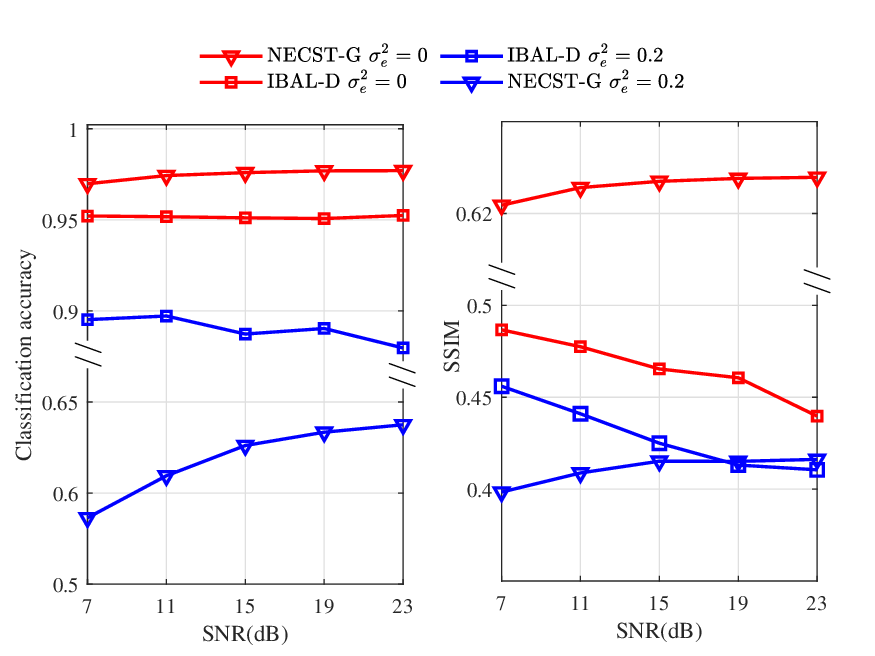}
\caption{Performance over rapidly changing channel conditions.}
    \label{kuaibian}
\end{figure}
To show the performance of the proposed IBAL approach over rapidly changing channel conditions, we further conduct more experiments.
Specifically, when the channel is rapidly changing, the estimated channel information $\hat{h}$ may be outdated.
We model the fast variation of the channel as a Gaussian random variable, and the real channel can be expressed as
\begin{equation}
h=\hat{h}+h_{\epsilon},
\end{equation}
where $h_{\epsilon}$ is Gaussian distributed with zero mean and covariance $\sigma^2_e$.
Simulation results are demonstrated in Fig. \ref{kuaibian}.
It is shown that the comparison algorithm can achieve better privacy protection performance due to the channel mismatch, resulting in a lower SSIM. However, its classification accuracy also significantly decreases due to the channel mismatch. On the contrary, our proposed algorithm possesses better robustness and significantly outperforms the comparison algorithms in terms of classification accuracy under fast-varying channels. Moreover, our method can realize the improvement of privacy protection performance with the increase of SNR due to the introduction of the adaptive weight adjustment strategy.

\subsection{Test with Imperfect Source Data}
\begin{figure}[t]
    \centering
\includegraphics[width=0.95\linewidth]{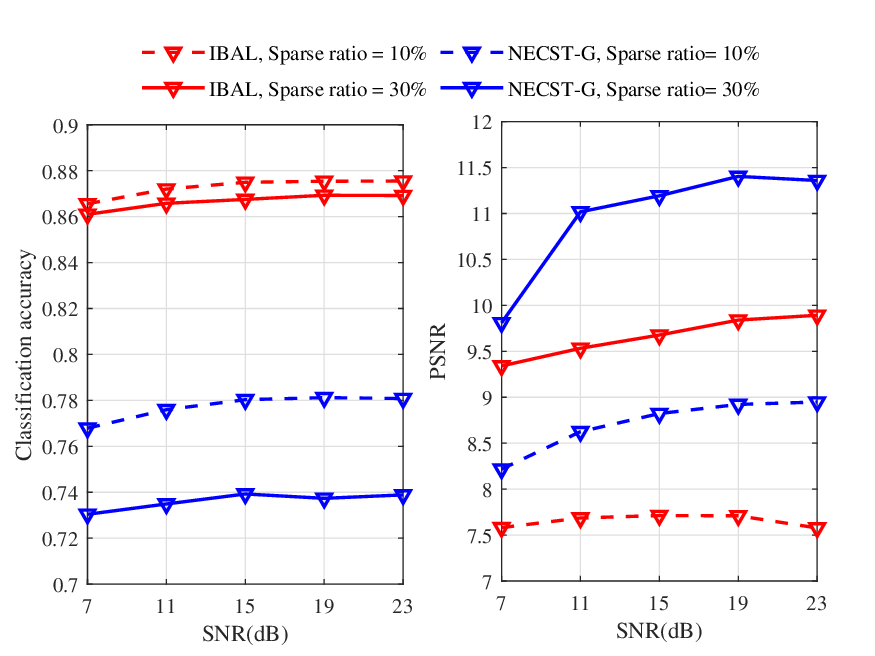}
\caption{Test results with imperfect source data of different sparsity ratios.}
    \label{xishu}
\end{figure}

The quality of the data also affects the performance of the training. To show the robustness of the proposed IBAL approach with an imperfect source
data, we conduct experiments on sparse source data with different sparsity ratios.
Specifically, we randomly zero out some elements in the data set to achieve sparsification. In Fig.~\ref{xishu} we perform sparse processing on the CIFAR-$10$ dataset with sparse ratios of $10\%$ and $30\%$, respectively. It can be observed that as the sparsity increases, both our proposed algorithm and the comparative algorithm experience a decrease in classification accuracy and SSIM performance. This is because data sparsity leads to a reduction in useful information, which in turn affects the training performance. However, despite this decline, our algorithm outperforms the comparative algorithm in terms of both classification accuracy and privacy protection. Specifically, the classification accuracy of the comparative algorithm significantly decreases due to the impact of data sparsity, whereas our proposed algorithm utilizes information bottleneck to better extract meaningful semantic information. As a result, it demonstrates robust performance even in sparse data scenarios.

\subsection{Test with Large-scale Dataset}

\begin{figure}[t]
    \centering
\includegraphics[width=0.95\linewidth]{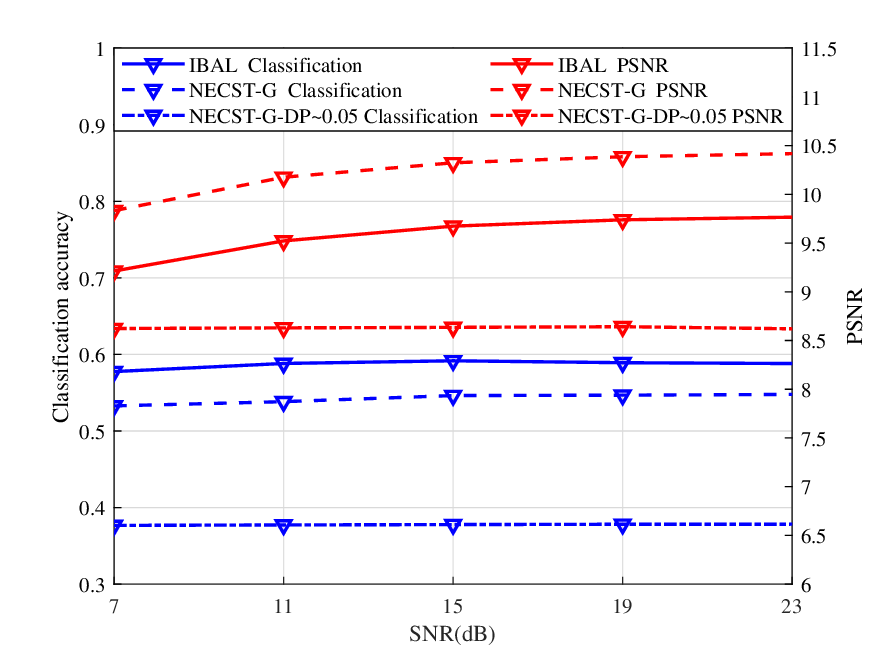}
\caption{Classification accuracy and SSIM of IBAL and NECST-G and NECST-G$\sim$0.05 on the ImageNet2012 dataset.}
    \label{bigdata}
\end{figure}
In order to enhance the generality of our proposed method across multiple datasets, we introduce a data adaptation network that maps large-scale complex datasets into lower-dimensional and easier-to-process datasets. To evaluate the performance of the proposed IBAL approach with large-scale dataset, we conduct more experiments on ImageNet2012 dataset.
ImageNet2012 contains $1000$ image classes, the training dataset has about $1300$ images per class, and the validation set has $50$ images per class.
For the experiments in this subsection, we select $10$ of these classes.
Following the basic settings in Section IV.B, we compare the proposed IBAL with NECST-G and NECST-G-DP$\sim0.05$.
Fig. \ref{bigdata}. shows that our proposed method has the best classification accuracy and  good privacy protection.

\begin{figure}[t]
    \centering
    \includegraphics[width=0.95\linewidth]{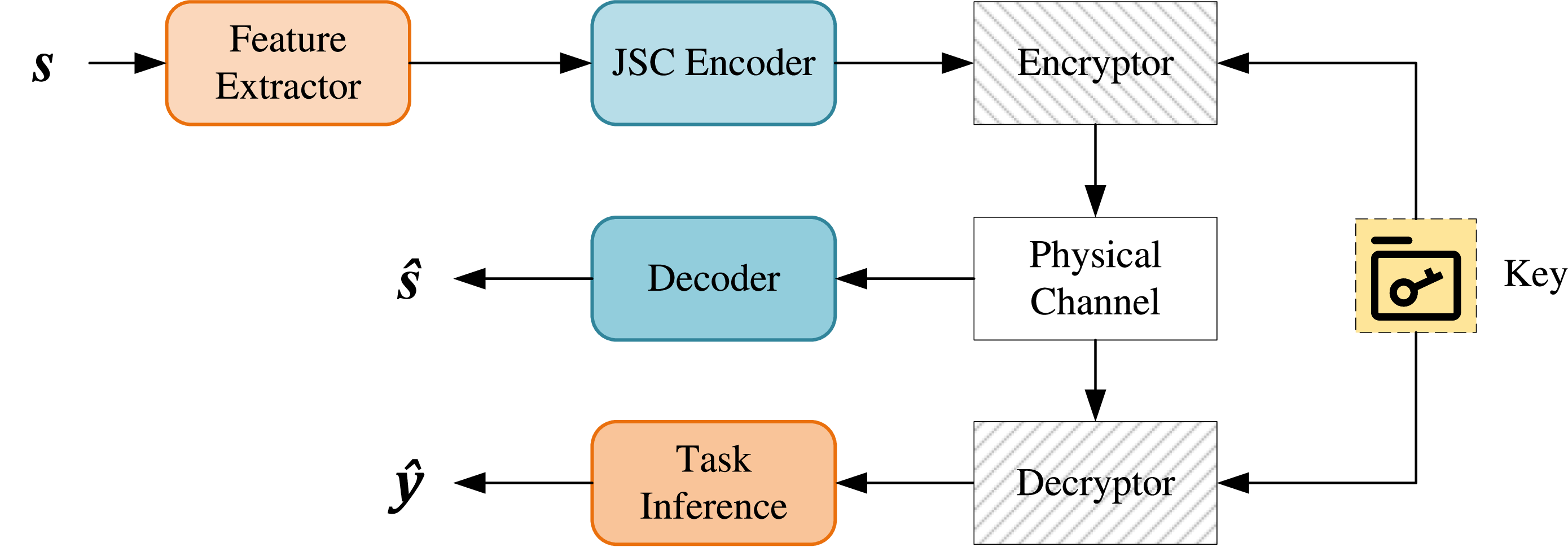}
    \caption{The framework of combining IBAL with encryption.}
    \label{IBE}
\end{figure}

\begin{figure}
    \centering
\includegraphics[width=0.9\linewidth]{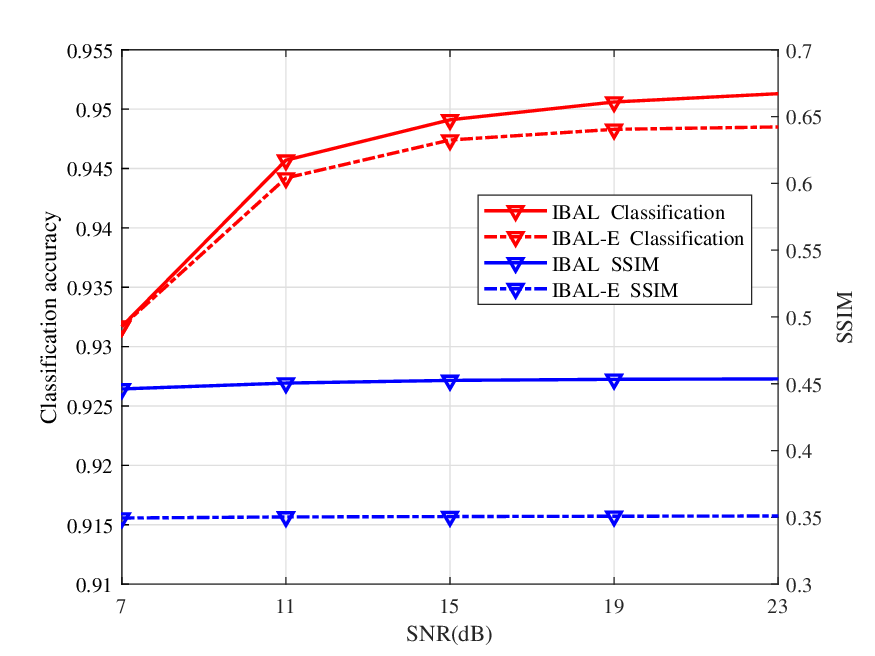}
    \caption{Classification accuracy and SSIM  of combining IBAL with encryption on the MNIST dataset.}
    \label{fig:enter-label}
\end{figure}


\subsection{Experiments of Combining IBAL with Encryption}
The proposed IBAL approach and existing privacy-preserving techniques are not mutually exclusive. In Fig. \ref{IBE}, we give the framework of combining IBAL with a non-blackbox approach. Specifically, an encryptor is incorporated to perform encryption using the key after the output of the JSC encoder, and a decryptor is introduced before the task inference network using the shared key, and the key is only shared between the transmitter and receiver, but cannot be accessed by the decoder. Then
in Fig. \ref{fig:enter-label}, we present the performance of the proposed algorithm in combination with the non-blackbox approach, which is depicted by the `IBAL-E' curve in Fig. \ref{fig:enter-label}. It is shown that integrating non-black-box algorithms with the proposed approach can substantially enhance the performance of privacy protection. Specifically, based on the SSIM metric, the performance shows a significant improvement of $21\%$. At the same time, IBAL-E still maintains high classification accuracy.


\section{Conclusion}

In this paper, we have proposed a privacy-preserving task-oriented semantic communication system, named IBAL, which could effectively protect users' privacy while enabling low-latency edge inference.
In IBAL, task-related information is extracted from the input at the transmitter based on the IB theory.
To overcome the computational difficulty in calculating some mutual information metrics needed for the IB loss,
we have derived a tractable variational upper bound for the IB loss by leveraging the variational approximation.
Moreover, we have adopted adversarial learning to make it more difficult to uncover the original data from the intercepted transmitted features through model inversion.
Extensive experimental results have demonstrated that the IBAL achieves a significantly better privacy-utility trade-off.
Further, we have extended the IBAL to make it more suitable under dynamic channel conditions.
In particular, to avoid manually tuning the loss weights,
we have treated the privacy-preserving task-oriented semantic communication problem as a multi-objective optimization problem
and solved it with the MGDA. Extensive simulation results show that our proposed algorithm is effective on dynamic channels, fast-varying channels, imperfect datasets, and large-scale datasets.

\begin{appendices}
\vspace{-0.2cm}
\section{Derivation of Variational Upper Bound for $\mathcal{L}_{IB}$}

In this appendix, we derive the variational upper bound of $\mathcal{L}_{IB}$, i.e., $\mathcal{\overline{L}}_{VIB}$.
The process is divided into the derivation of the variational upper bound of $I(\hat{Z};S)$
and the derivation of the variational lower bound of $I(\hat{Z};Y)$, as shown below.

The definition of $I(\hat{Z};S)$ can be written as
\begin{equation}
\begin{split}
\label{IZS}
I(\hat{Z};S)&=\int p(s)p_{\theta}(\hat{z}|s)\log \frac{p_{\theta}(\hat{z}|s)}{p(\hat{z})}  \,{dsd\hat{z}},
\end{split}
\end{equation}
where $p(\hat{z})$ is intractable, as described in Section III.
We introduce the variational approximation $q(\hat{z})$ into \eqref{IZS},
so that \eqref{IZS} can be rewritten as
\begin{equation}
\begin{split}
\label{izszk}
I(\hat{Z};S)&=\int p(s)p_{\theta}(\hat{z}|s) \log \frac{p_{\theta}(\hat{z}|s)q(\hat{z})}{p(\hat{z})q(\hat{z})}   \,{dsd\hat{z}}\\
&=\int p(s)p_{\theta}(\hat{z}|s)\log \frac{p_{\theta}(\hat{z}|s)}{q(\hat{z})} \,{dsd\hat{z}} \\&\quad+ \int p(\hat{z})\log \frac{q(\hat{z})}{p(\hat{z})} \,{d\hat{z}}\\
&=\int p(s)p_{\theta}(\hat{z}|s)\log \frac{p_{\theta}(\hat{z}|s)}{q(\hat{z})}   \,{dsd\hat{z}} \\&\quad -KL(p(\hat{z})||q(\hat{z})),
\end{split}
\end{equation}
where KL divergence is non-negative. Drop the term and we obtain the variational upper bound of $I(\hat{Z};S)$, i.e.,
\begin{equation}
\begin{split}
\label{izsub}
I(\hat{Z};S)&\leq \int p(s)p_{\theta}(\hat{z}|s) \log \frac{p_{\theta}(\hat{z}|s)}{q(\hat{z})} \,{dsd\hat{z}}\\
&= \int p(s,y)p_{\theta}(\hat{z}|s)  \log \frac{p_{\theta}(\hat{z}|s)}{q(\hat{z})}   \,{dsdyd\hat{z}}\\
&= \mathbb{E}_{p(s,y)} KL(p_{\theta}(\hat{z}|s)||q(\hat{z})).
\end{split}
\end{equation}

$I(\hat{Z};Y)$ can be written as
\begin{equation}
\begin{split}
\label{izy}
I(\hat{Z};Y)&= \int p(\hat{z})p(y|\hat{z}) \log \frac{p(y|\hat{z})}{p(y)}  \,{dyd\hat{z}}\\
&= \int p(\hat{z})p(y|\hat{z})  \log p(y|\hat{z})   \,{dyd\hat{z}}\\&\quad- \int p(y)\log p(y) \,{dy} \\
&= \int p(\hat{z})p(y|\hat{z})  \log p(y|\hat{z})   \,{dyd\hat{z}} +H(Y),
\end{split}
\end{equation}
where $H(Y)$, representing the entropy of $Y$, is a constant.
Drop $H(Y)$ and we get a lower bound of $I(\hat{Z};Y)$ as follows:
\begin{equation}
\begin{split}
\label{izylb}
I(\hat{Z};Y)&\geq \int p(\hat{z})p(y|\hat{z})  \log p(y|\hat{z}) \,{dyd\hat{z}} \\
&= \int p(s,y,\hat{z})  \log p(y|\hat{z}) \,{dsdyd\hat{z}}\\
&= \int p(s,y)p_{\theta}(\hat{z}|s)   \log p(y|\hat{z}) \,{dsdyd\hat{z}}\\
&= \mathbb{E}_{p(s,y)}\mathbb{E}_{p_{\theta}(\hat{z}|s)}  [\log p(y|\hat{z})].
\end{split}
\end{equation}
Since $p(y|\hat{z})$ is intractable, we use $q_{\phi}(y|\hat{z})$ to approximate $p(y|\hat{z})$,
so that \eqref{izylb} can rewritten as
\begin{equation}
\begin{split}
\label{izyvlb}
I(\hat{Z};Y)&\geq \mathbb{E}_{p(s,y)}\mathbb{E}_{p_{\theta}(\hat{z}|s)}  [\log q_{\phi}(y|\hat{z})].
\end{split}
\end{equation}

Combining \eqref{izsub} and \eqref{izyvlb}, we obtain the tractable variational upper bound of $\mathcal{L}_{IB}$ as follows:
\begin{equation}
\begin{split}
\label{ibvub}
\mathcal{\overline{L}}_{VIB} &=\mathbb{E}_{p(s,y)}\{\mathbb{E}_{p_{\theta}(\hat{z}|s)}[- \log q_{\phi}(y|\hat{z})] \\&\quad+\beta KL(p_{\theta}(\hat{z}|s)||q(\hat{z}))\}.
\end{split}
\end{equation}

\section{The Solution of \eqref{obj5}}

For simplicity, we use $\boldsymbol{\Theta}$ and $\boldsymbol{\overline{\Theta}}$ to represent $\bigtriangledown_{\theta} \mathcal{\overline{L}}_{VIB}$ and $-\bigtriangledown_{\theta}(\frac{1}{1+\sigma^2} \mathcal{L}_{MSE})$ in \eqref{obj5}, respectively,
and then \eqref{obj5} is rewritten as
\begin{equation}
\begin{split}
\label{obj6}
\min_{\lambda \in[0,1]}\parallel \lambda \boldsymbol{\Theta} + (1-\lambda)\boldsymbol{\overline{\Theta}} \parallel_2^2.
\end{split}
\end{equation}

From \cite{3326943},
the optimization problem defined by \eqref{obj6} is equivalent to finding the minimum norm point in the convex hull of two points.
In computational geometry, it is equivalent to finding the closest point in the convex hull to a given query point.
Following the work\cite{3326943}, we use the Frank-Wolfe algorithm developed in \cite{3042817} to solve our optimization problem at hand, i.e., \eqref{obj6}.
The specific procedures for Frank-Wolfe are illustrated in Algorithm \ref{FWS}.

\renewcommand{\algorithmicrequire}{\textbf{Initialization:}}
\begin{algorithm} [htb]
\caption{$\min_{\lambda\in[0,1]}\parallel \lambda \boldsymbol{\Theta} + (1-\lambda)\boldsymbol{\overline{\Theta}} \parallel_2^2$}
\label{minalgorithm}
    \begin{algorithmic}[1]
        \STATE \textbf{if $\boldsymbol{\Theta^\top \overline{\Theta} \geq \Theta^\top \Theta}$ then}
        \STATE \quad $\lambda=1$.
        \STATE \textbf{else if $\boldsymbol{\Theta^\top \overline{\Theta} \geq \overline{\Theta}^\top \overline{\Theta}}$ then}
        \STATE \quad $\lambda=0$.
        \STATE \textbf{else}
        \STATE \quad $\lambda=\frac{(\boldsymbol{\overline{\Theta}}-\boldsymbol{\Theta})^\top \boldsymbol{\overline{\Theta}}}{\parallel \boldsymbol{\Theta}-\boldsymbol{\overline{\Theta}} \parallel^2_2}$.
        \STATE \textbf{end if}
     \end{algorithmic}
\end{algorithm}

\renewcommand{\algorithmicrequire}{\textbf{Initialization:}}
\begin{algorithm} [htb]
\caption{Procedures for Frank-Wolfe Algorithm}
\label{FWS}
    \begin{algorithmic}[1]
        \STATE  Initialize $\boldsymbol{\Lambda}=(\lambda,1-\lambda)=(0.5, 0.5)$.
        \STATE  Precompute $\rm\textbf{M}$,\\
        \quad $\rm\textbf{M}=\left[
        \begin{array}{lr}
        \Theta^\top \Theta, \quad \Theta^\top\overline{\Theta}&\\
        \overline{\Theta}^\top \Theta, \quad  \overline{\Theta}^\top \overline{\Theta}&
        \end{array}\right]$\\
        \STATE  \textbf{repeat}
        \STATE \quad $\hat{t}= \arg \min_\emph{r} \sum_\emph{t} \boldsymbol{\Lambda}_t\textbf{M}_{rt}$.
        \STATE  \quad Using Algorithm \ref{minalgorithm}: \\
        \quad $\hat{\lambda}= \arg \min_{\lambda}((1-\lambda)\boldsymbol{\Lambda}+\lambda\boldsymbol{e_{\hat{t}}})^\top\textbf{M}((1-\lambda)\boldsymbol{\Lambda}+\lambda\boldsymbol{e_{\hat{t}}})$.
        \STATE \quad $\boldsymbol{\Lambda}=(1-\hat{\lambda})\boldsymbol{\Lambda} + \hat{\lambda}\boldsymbol{e_{\hat{t}}}$.
        \STATE  \textbf{until} $\hat{\lambda} \sim 0$ or number of iterations limit.
        \STATE \textbf{return} $\boldsymbol{\Lambda}$.
     \end{algorithmic}
\end{algorithm}

\section{Proof of proposition 1}

\begin{IEEEproof}
To facilitate the understanding of the proof, we first provide the iteration formula of variables for Adam, i.e.
\begin{equation}
\label{vif}
\left\{
        \begin{array}{lr}
        \mathbf {m}^{(t)}=\beta_1 \mathbf {m} ^{(t-1)}+ (1-\beta_1) \mathbf {g}_t,&\\
        \hat{\mathbf {m}}^{(t)}= \mathbf {m} ^{(t)} / (1-\beta_1^t),&\\
        \mathbf {v}^{(t)}=\beta_2 \mathbf {v} ^{(t-1)}+ (1-\beta_2)\mathbf {g}^2_t,&\\
        \hat{\mathbf {v}}^{(t)}= \max \left( \hat{\mathbf {v}}^{(t-1)}, \mathbf {v}^{(t)} / \left( 1-\beta_2^t \right)\right),&\\
        \textbf{\emph{w}}^{(t+1)} = \textbf{\emph{w}}^{(t)} - \eta_t \hat{\mathbf {m}}^{(t)} / \sqrt{\hat{\mathbf {v}}^{(t)}},&\\
        \end{array}
\right.
\end{equation}
where $ \mathbf {g}_t = \nabla F_t \left(\textbf{\emph{w}}^{(t)}\right)$.

According to the analysis in \cite{Chen2019iclr}, we can get
\begin{equation}
\begin{split}
   \label{EUB}
    &\mathbb{E} \left[  \sum_{t=1}^T \eta_t \left\langle \nabla F \left(\textbf{\emph{w}}^{(t)}\right), \nabla F \left(\textbf{\emph{w}}^{(t)}\right) / \sqrt{\hat{\mathbf {v}}^{(t-1)}} \right\rangle \right]\\
    &\leq \mathbb{E} \Bigg[ \underbrace{ C_1 \sum_{t=1}^T \left\| \eta_t \mathbf {g}_t / \sqrt{\hat{\mathbf {v}}^{(t)}} \right\|^2}_{\rm {Term \ A}} \\&\quad+ \underbrace{C_2 \sum_{t=2}^T \left\| \frac{\eta_t}{\sqrt{\hat{\mathbf {v}}^{(t)}}} - \frac{\eta_{t-1}}{\sqrt{\hat{\mathbf {v}}^{(t-1)}}} \right\|_1}_{\rm {Term \ B}} \\&\quad+ \underbrace{C_3 \sum_{t=2}^{T-1} \left\| \frac{\eta_t}{\sqrt{\hat{\mathbf {v}}^{(t)}}} - \frac{\eta_{t-1}}{\sqrt{\hat{\mathbf {v}}^{(t-1)}}} \right\|^2}_{ \rm {Term \ C} } \Bigg] + C_4,
\end{split}
\end{equation}
where $C_1$, $C_2$, $C_3$ are constants independent of the vector dimension $d$ and the number of iterations $T$ , $C_4$ is a constant independent of $T$.
Next, we will handle items: $\rm {Term \ A}$, $\rm {Term \ B}$, and $\rm {Term \ C}$, separately.

For $\rm {Term \ A}$, $\hat{\mathbf {v}}^{(1)}$ satisfies: ${\rm min}_{i=1,2,\ldots,d} \left( \sqrt{ \hat{v}^{(1)}_i} \right) \geq c > 0$.
Proof by contradiction is easy to demonstrate. If the condition is not satisfied, i.e., $ \exists \sqrt{ \hat{v}^{(1)}_i} = 0$,
then $\textbf{\emph{w}}$ cannot be updated in \eqref{vif}, and the algorithm will cause an error.
Substituting the condition in $\rm {Term \ A}$, we can obtain
\begin{equation}
\begin{split}
   \label{terma}
    &\mathbb{E} \left[ \sum_{t=1}^T \left\| \eta_t \mathbf {g}_t / \sqrt{\hat{\mathbf {v}}^{(t)}} \right\|^2 \right] \\&\overset{(a)}{\leq}  \mathbb{E} \left[ \sum_{t=1}^T \left\| \eta_t \mathbf {g}_t / c \right\|^2 \right]
   \\&=  \mathbb{E} \left[ \sum_{t=1}^T \left\| \frac{1}{\sqrt{t}} \mathbf {g}_t / c \right\|^2 \right]\\
     &=  \mathbb{E} \left[ \sum_{t=1}^T \left( \frac{1}{c\sqrt{t}} \right)^2  \left\|  \mathbf {g}_t \right\|^2 \right]
    \\&\overset{(b)}{\leq} G^2 / c^2 \sum_{t=1}^T \frac{1}{t} 
    \\& \overset{(c)}{\leq} G^2 / c^2 (1+\log T),
\end{split}
\end{equation}
where (a) is due to $\hat{v}^{(t)}_i = \max \left( \hat{v}^{(t-1)}_i, \frac{v^{(t)}}{1-\beta_2^t} \right) \geq  \hat{v}^{(t-1)}_i$,
(b) is due to Assumption 4, i.e., $\left \| \mathbf {g}_t \right \| = \left \| \nabla F_t\left(\textbf{\emph{w}}^{(t)}\right) \right \| \leq G$,
and (c) is due to $\sum_{t=1}^T \frac{1}{t} \leq  1+\log T$.

For $\rm {Term \ B}$, we have
\begin{equation}
\small
\begin{split}
   \label{termb}
  &\mathbb{E} \left[ \sum_{t=2}^T \left\| \frac{\eta_t}{\sqrt{\hat{\mathbf {v}}^{(t)}}} - \frac{\eta_{t-1}}{\sqrt{\hat{\mathbf {v}}^{(t-1)}}} \right\|_1 \right]
    \\& \overset{(a)}{=} \mathbb{E}  \left[\sum_{i=1}^d \sum_{t=2}^T \left( \frac{\eta_{t-1}}{\sqrt{\hat{v}^{(t-1)}_i}} - \frac{\eta_{t}}{\sqrt{\hat{v}^{(t)}_i}} \right) \right] \\
    & \overset{(b)}{=} \mathbb{E}  \left[\sum_{i=1}^d \left( \frac{\eta_{1}}{\sqrt{\hat{v}^{(1)}_i}} - \frac{\eta_{T}}{\sqrt{\hat{v}^{(T)}_i}} \right) \right] \\&\leq \mathbb{E}  \left[ \sum_{i=1}^d  \frac{\eta_{1}}{\sqrt{\hat{v}^{(1)}_i}} \right] \\&\leq d / c,
\end{split}
\end{equation}
where (a) is because $\hat{v}^{(t)}_i \geq \hat{v}^{(t-1)}_i$ and $\eta_{t} \leq \eta_{t-1}$,
and (b) is by expanding $\sum_{t=2}^T$ and  merging.

For $\rm {Term \ C}$, it can be transformed into
\begin{equation}
\begin{split}
   \label{termc}
    &\mathbb{E} \left[ \sum_{t=2}^{T-1} \left\| \frac{\eta_t}{\sqrt{\hat{\mathbf {v}}^{(t)}}} - \frac{\eta_{t-1}}{\sqrt{\hat{\mathbf {v}}^{(t-1)}}} \right\|^2 \right] 
    \\&\overset{(a)}{\leq} \mathbb{E} \left[ \frac{1}{c} \sum_{t=2}^{T-1} \left\| \frac{\eta_t}{\sqrt{\hat{\mathbf {v}}^{(t)}}} - \frac{\eta_{t-1}}{\sqrt{\hat{\mathbf {v}}^{(t-1)}}} \right\|_1 \right]
    \\&\leq d / c^2,
\end{split}
\end{equation}
where (a) is due to $ \left\| \frac{\eta_t}{\sqrt{\hat{\mathbf {v}}^{(t)}}} - \frac{\eta_{t-1}}{\sqrt{\hat{\mathbf {v}}^{(t-1)}}}  \right\|_1 \leq 1/c$ which can be seen from \eqref{termb}.

Combining with \eqref{EUB}, \eqref{terma}, \eqref{termb}, and \eqref{termc}, we obtain
\begin{equation}
\small
\begin{split}
   \label{eq42}
    & \mathbb{E} \Bigg[  C_1 \sum_{t=1}^T \left\| \eta_t \mathbf {g}_t / \sqrt{\hat{\mathbf {v}}^{(t)}} \right\|^2 + C_2 \sum_{t=2}^T \left\| \frac{\eta_t}{\sqrt{\hat{\mathbf {v}}^{(t)}}} - \frac{\eta_{t-1}}{\sqrt{\hat{\mathbf {v}}^{(t-1)}}} \right\|_1   \\&\quad+ C_3 \sum_{t=2}^{T-1} \left\| \frac{\eta_t}{\sqrt{\hat{\mathbf {v}}^{(t)}}} - \frac{\eta_{t-1}}{\sqrt{\hat{\mathbf {v}}^{(t-1)}}} \right\|^2 \Bigg] + C_4\\
     & \leq C_1 G^2 /c^2 \left(1 + \log T\right) + C_2 d / c + C_3d/c^2 + C_4.
\end{split}
\end{equation}

Scaling the left side of \eqref{EUB}, we get
\begin{equation}
\begin{split}
   \label{eq43}
    &\mathbb{E} \left[  \sum_{t=1}^T \eta_t \left\langle \nabla F \left(\textbf{\emph{w}}^{(t)}\right), \nabla F \left(\textbf{\emph{w}}^{(t)}\right) / \sqrt{\hat{\mathbf {v}}^{(t-1)}} \right\rangle \right] \\&\overset{(a)}{\geq} \mathbb{E} \left[ \sum_{t=1}^T \frac{1}{G \sqrt{t}} \left\| \nabla F \left(\textbf{\emph{w}}^{(t)}\right) \right\|^2 \right]\\
    & \geq \frac{\sqrt{T}}{G} {\rm min}_{t=1,2,\ldots,T} \mathbb{E} \left[ \left\| \nabla F \left(\textbf{\emph{w}}^{(t)}\right) \right\|^2 \right].
\end{split}
\end{equation}
As $\hat{\mathbf {v}}^{(t)}$ is the exponential moving average of $\mathbf {g}_t^2$, and $ \left\|\mathbf {g}_t\right\| \leq G$,
we obtain $\hat{\mathbf {v}}^{(t)} \leq G^2 $.
And (a) is precisely due to this fact.

From \eqref{EUB}, \eqref{eq42}, and \eqref{eq43}, we derive
\begin{equation}
\begin{split}
   \label{eq44}
    & \frac{\sqrt{T}}{G} {\rm min}_{t=1,2,\ldots,T} \mathbb{E} \left[ \left\| \nabla F \left(\textbf{\emph{w}}^{(t)}\right) \right\|^2 \right]  \\&\leq C_1 G^2 /c^2 \left(1 + \log T\right) + C_2 d / c + C_3d/c^2 + C_4.
\end{split}
\end{equation}

Rearranging \eqref{eq44}, we get
\begin{equation}
\begin{split}
   \label{eq45}
    & {\rm min}_{t=1,2,\ldots,T} \mathbb{E} \left[ \left\| \nabla F \left(\textbf{\emph{w}}^{(t)}\right) \right\|^2 \right]\\
    & \leq \frac{G}{\sqrt{T}} \left(C_1 G^2 /c^2 \left(1 + \log T\right) + C_2 d / c + C_3d/c^2 + C_4\right) \\
    & = \frac{1}{\sqrt{T}} \left(G_1+G_2 \log T\right),
\end{split}
\end{equation}
where $G_1$ and $G_2$ are constants independent of $T$.
Then, the proof is completed.
\end{IEEEproof}

\end{appendices}

\ifCLASSOPTIONcaptionsoff
  \newpage
\fi

\bibliography{reference}
\vspace{-25 pt}

\end{document}